\documentclass[a4paper, 11pt]{article}
\usepackage[utf8]{inputenc}
\usepackage[T1]{fontenc}
\usepackage{graphicx}
\usepackage{longtable}
\usepackage{wrapfig}
\usepackage{rotating}
\usepackage[normalem]{ulem}
\usepackage{amsmath}
\usepackage{amssymb}
\usepackage{capt-of}
\usepackage{hyperref}
\usepackage[american]{babel}
\usepackage{amssymb}
\usepackage{amsmath}
\usepackage{upgreek}
\usepackage{amsfonts}
\usepackage{graphicx}
\usepackage{epstopdf}
\usepackage[]{enumerate}
\usepackage{algorithm}
\usepackage{algpseudocode}
\usepackage{amsthm}
\usepackage{mathrsfs}
\usepackage{csquotes}
\usepackage{cleveref}
\usepackage{bbm}
\usepackage{tikz}
\usetikzlibrary{shapes.misc, arrows}
\usepackage{subcaption}
\usepackage{wasysym}
\usetikzlibrary{decorations.text, fit}
\usepackage{aligned-overset}
\crefname{prop}{property}{properties}
\crefname{Assumption}{assumption}{assumptions}
\crefalias{subassumption}{Assumption}
\crefalias{enumi}{prop}
\crefname{equation}{}{}
\usepackage[shortlabels]{enumitem}
\setlist[enumerate]{leftmargin=.5in}
\setlist[itemize]{leftmargin=.5in}
\newtheorem{Theorem}{Theorem}[section]
\newtheorem{Lemma}[Theorem]{Lemma}
\newtheorem{Proposition}[Theorem]{Proposition}

\newtheorem{Example}[Theorem]{Example}
\newtheorem{Remark}[Theorem]{Remark}

\numberwithin{equation}{section}

\newcounter{subassumption}[subsection]

\makeatletter
\newcommand{\leqnomode}{\tagsleft@true\let\veqno\@@leqno}
\newcommand{\reqnomode}{\tagsleft@false\let\veqno\@@eqno}
\renewcommand{\p@subassumption}{\theAssumption}
\makeatother

\usepackage{booktabs}

\usepackage{float}
\newcommand{\email}[1]{\protect\href{mailto:#1}{#1}}

\usepackage[autocite=plain, backend=bibtex, doi=true, url=true, hyperref=true, maxcitenames=2, giveninits=true]{biblatex}
\renewbibmacro{in:}{%
\ifentrytype{article}
{}
{\bibstring{in}%
\printunit{\intitlepunct}}}
\AtEveryBibitem{%
\clearfield{issn} 
\clearfield{doi} 
\ifentrytype{online}{}{
\clearfield{url}
}
}
\DeclareFieldFormat*{title}{#1}
\DeclareFieldFormat{journaltitle}{#1\isdot}
\DeclareFieldFormat*{title}{\mkbibitalic{#1}}
\usepackage[binary-units=true]{siunitx}
\DeclareSIUnit\px{px}
\usepackage{textcomp}
\usepackage{import}
\usepackage{xifthen}
\usepackage{pdfpages}
\usepackage{transparent}

\newbox\namebox
\newdimen\signboxdim
\def\signature#1{%
\setbox\namebox=\hbox{#1}
\signboxdim=\dimexpr(\wd\namebox+1cm)
\parbox[t]{\signboxdim}{%
\centering
\hrulefill\\    
#1
\par}%
}
\author{Anna Aksamit\footnotemark[2] \and Ivan Guo\footnotemark[3]\ \footnotemark[4] \and Shidan Liu\footnotemark[3] \and Zhou Zhou\footnotemark[2]}
\tikzset{%
highlight/.style={rectangle, rounded corners, fill = red!15, draw, fill opacity=0.25, dashed, minimum width = 0.01, inner sep = -2}
}
\tikzset{%
highlight2/.style={rectangle, rounded corners, fill = blue!15, draw, fill opacity=0.25, dashed, minimum width = 0.01, inner sep = -2}
}
\tikzstyle{line} = [draw, -latex']
\date{}
\title{Superhedging duality for multi-action options under model uncertainty with information delay}
\hypersetup{
 pdfauthor={Shidan Liu},
 pdftitle={Superhedging duality for multi-action options under model uncertainty with information delay},
 pdfkeywords={},
 pdfsubject={},
 pdfcreator={Emacs 30.0.50 (Org mode 9.6.6)}, 
 pdflang={English}}
\begin{document}

\maketitle

\renewcommand{\thefootnote}{\fnsymbol{footnote}}
\footnotetext[2]{School of Mathematics and Statistics, The University of Sydney, Sydney, NSW, Australia 
(\email{anna.aksamit@sydney.edu.au}, \email{zhou.zhou@sydney.edu.au}). The research of A. Aksamit was supported by the Australian Research Council DECRA Fellowship DE200100896.}
\footnotetext[3]{School of Mathematics, Monash University, Clayton, VIC, Australia
(\email{Ivan.Guo@monash.edu}, \email{Shidan.Liu@monash.edu}).}
\footnotetext[4]{Center for Quantitative Finance and Investment Strategies, Monash University, Clayton, VIC, Australia}
\footnotetext{The authors thank Marek Rutkowski for helpful discussions and comments.}
\renewcommand{\thefootnote}{\arabic{footnote}}

\begin{abstract}
We consider the superhedging price of an exotic option under nondominated model uncertainty in discrete time in which the option buyer chooses some action from an (uncountable) action space at each time step. 
By introducing an enlarged space we reformulate the superhedging problem for such an exotic option as a problem for a European option, which enables us to prove the pricing-hedging duality. Next, we present a duality result that, when the option buyer’s action is observed by the seller up to \(l\) periods later, the superhedging price equals the model-based price where the option buyer has the power to look into the future for \(l\) periods.
\end{abstract}


\section{Introduction}
\label{sec:org3d29d62}
The robust approach to pricing and hedging of financial derivatives addresses the traditional approach's inability to account for model misspecification risk, by considering a collection of all possible models which are represented by probability measures, as opposed to pre-defining a probability measure that we guess to be the correct one in the traditional approach.
Superhedging is a trading strategy whose value should be able to cover the payoff at all possible states and times.
The duality between the minimum superhedging price and the supremum of expectations of the payoff (the model price) is a goal of much of the literature in robust finance, 
as it proves the optimality of both prices.
Duality under volatility uncertainty in continuous time was studied in, for example, Denis and Martini \cite{denis_2006_theor_frame}, Nutz and Neufeld \cite{neufeld_2013_super_volat}, Soner et al. \cite{soner_2013_dual_targe}, through different approaches.

Hobson \cite{hobson_1998_robus_lookb} initiated the approach of superhedging using dynamic strategies for the underlying risky assets and static strategies for a given set of liquid options.
This approach is desirable in the robust framework, as one often wants to include additional marketed securities for trading.  
In discrete time, Bouchard and Nut \cite{bouchard_2015_arbit_nondo} introduced the notion of no arbitrage in a quasi sure sense, and showed a general pricing-hedging duality with the presence of a finite number of liquid options.
Burzoni et al. \cite{burzoni_2017_model_free} obtained a similar duality result, using a path-wise approach.
General pricing-hedging duality in continuous time is shown in, for example, Biagini et al. \cite{biagini_2017_robus_funda}, Dolinsky and Soner \cite{dolinsky_2014_marti_optim}, Hou and Ob\l{\'o}j \cite{hou_2018_robus_conti}.

Apart from works that are mainly focused on the robust superhedging problem of European options, there are various works devoted to American options. 
Neuberger \cite{neuberger_2007_bound_ameri} showed that the superhedging price for an American option may be strictly larger than the highest model price.
Bayraktar et al. \cite{bayraktar_2013_hedgi_uncer} established sub and superhedging dualities for American options in the setup of \cite{bouchard_2015_arbit_nondo}.
In addition, it is shown in \cite{bayraktar_2013_hedgi_uncer} that with the presence of European options, the order of the infimum and the supremum in the subhedging dual representation may not be changed due to the lack of dynamic programming principle for the underlying martingale measure set. Several alternative definitions of superhedging prices were also discussed and compared in \cite{bayraktar_2013_hedgi_uncer}.
In a discrete time, discrete state space market with statically traded European vanilla options, Hobson and Neuberger \cite{hobson_2017_model_uncer} obtained superhedging duality by adopting a weak formulation.
That is, the optimization runs over filtered probability spaces supporting a martingale rather than over martingale measures on a fixed filtered path space, which allows richer information structures and hence more stopping times.
Soon after, Bayraktar and Zhou \cite{bayraktar_2017_hedgi_ameri} improved the duality result in \cite{hobson_2017_model_uncer}, and showed that in terms of model-based price, only randomized models, as in mixing the martingale measures in the original space, are relevant and other models proposed in \cite{hobson_2017_model_uncer} are redundant. After that, Aksamit et al. \cite{aksamit_2019_robus_prici} obtained a similar but more general duality result in the sense that their model also works in the martingale optimal transport setup of Beiglböck et al. \cite{beiglbock_2013_model_indep}.
To achieve the result, they enlarged the probability space, and the American options in the original space were converted to European options in the enlarged  space. They further gave insight that the duality gap as raised in \cite{neuberger_2007_bound_ameri} and \cite{bayraktar_2017_hedgi_ameri} is caused by the failure of the dynamic programming principle, and proposed to restore dynamic consistency and to restore duality, by considering fictitious extensions of the market where all assets can be traded dynamically.
Let us also mention the works Bayraktar and Zhou \cite{bayraktar_2016_utili_ameri} and \cite{bayraktar_2016_hedgi_liqui} where hedging American options are considered involving liquid American options.
In particular, they provided a unified framework and established sub and superhedging dualities for American options in a market consisting of dynamically traded stocks and statically traded (can be bought and sold) European and American options.
Another work related to superhedging American options is Herrmann and Stebegg \cite{herrmann_2019_robus_globe}, where a martingale optimal transport duality is established with given initial and terminal distributions, and it was shown that \emph{without any path restriction on the state space}, the superhedging prices of American, Asian, Bermudan, and European options with intermediate maturity are all the same.

Inspired by the above works, we aim to extend the arguments and duality results in \cite{aksamit_2019_robus_prici} as well as \cite{bouchard_2015_arbit_nondo} to account for the scenarios where the buyer of the option is allowed to choose some action from an action space, countable or uncountable, at each time step. 
In particular, by enlarging the probability space as that in \cite{aksamit_2019_robus_prici}   and applying analytic set theory techniques and measurable selection arguments, we are able to obtain a pricing-hedging duality for such exotic options. 
The main difficulty is to construct suitable analytic sets in order to apply measurable selection theorems so that we could prove the existence of an \(\epsilon\)-optimal sequence of actions by the option buyer and obtain a dynamic programming principle en route to the duality.
Furthermore, we also obtain a European style pricing-hedging duality in the enlarged space.

We further consider the scenario where the decisions made by the buyer is revealed to the seller with a variable delay periods that is capped by some number \(l\).
The delay increases the superhedging price and hence breaks the duality.
We show that the duality can be restored, by fictitiously allowing the buyer the capability to look into the future for \(l\) periods and make her decisions based on future information.

Our results generalize those in \cite{aksamit_2019_robus_prici,bayraktar_2016_hedgi_liqui,bayraktar_2017_hedgi_ameri,hobson_2017_model_uncer}, in the sense that in our setup the option buyer’s action space can be very general and uncountable. As a result, our work can be applied to the pricing of swing options on oil or natural gas which involves an uncountable action space for the option buyer (see \Cref{exmp:swing}). Another application is the superhedging of multiple American options possibly with certain constraint on exercise strategies, such as a restriction on the difference of consecutive exercise times, or the maximum amount of American options that can be exercised at each period (see \Cref{sec:exmp_multi}). Another contribution of our work lies in the result with information delay. Our duality provides an interesting financial observation: the \(l\)-period delay for the option seller in the primal hedging problem, to some extent, is equivalent to the \(l\)-period insider information for the option buyer in the dual problem.

The rest of this paper is organized as follows. 
\Cref{sec:setup} gives a quick recap on the model setup, formulates the space enlargement and reformulates the superhedging problem as one for a European option in the enlarged space.  \Cref{sec:main1} proves our first main result, i.e., the European pricing-hedging duality in the enlarged space (\Cref{thm:dual_bar}).
In \Cref{sec:info_delay}, we present a duality result as \Cref{thm:info_dual} in the delayed-information setup and present a delayed Snell envelope result.
More applicable examples of our model can be found in \Cref{sec:exmp}.

\section{Settings}
\label{sec:setup}
\subsection{Notation}
\label{sec:org797bdf9}
Given a measurable space \(\left( \Omega, \mathcal{F} \right)\), denote by \(\mathfrak{P}(\Omega)\) the set of all probability measures on \(\mathcal{F}\).
If \(\Omega\) is a topological space, then its Borel \(\sigma\)-algebra is denoted by \(\mathcal{B}(\Omega)\).
If \(\Omega\) is a Polish space, then a subset \(A \subseteq \Omega\) is \emph{analytic} if it is the image of a Borel subset of another Polish space under a Borel measurable mapping.
A function \(f: \Omega \to \overline{\mathbb{R}} := [-\infty, \infty]\) is \emph{upper semianalytic}, if the set \(\left\{ \omega \in \Omega: f(\omega) > c \right\}\) is analytic for all \(c \in \mathbb{R}\).
For a family \(\mathcal{P} \subseteq \mathfrak{P}(\Omega)\) of probability measures, a subset \(A \subseteq \Omega\) is called \(\mathcal{P}\)-\emph{polar} if \(A \subseteq A'\), 
for some \(A' \in \mathcal{F}\) satisfying \(\mathbb{P}(A') = 0\) for all \(\mathbb{P} \in \mathcal{P}\).
A property is said to hold \(\mathcal{P}\)-\emph{quasi surely} or \(\mathcal{P}\)-\emph{q.s.}, if it holds outside a \(\mathcal{P}\)-polar set. 
A set \(A \subseteq \Omega\) is universally measurable if it is measurable with respect to the universal completion of \(\mathcal{F}\), which is the \(\sigma\)-algebra \(\cap_{\mathbb{P} \in \mathfrak{P}(\Omega)} \mathcal{F}^{\mathbb{P}}\), with \(\mathcal{F}^{\mathbb{P}}\) being the \(\mathbb{P}\)-completion of \(\mathcal{F}\). 

\paragraph{A discrete-time model}
\label{sec:org73a2274}

Consider the discrete-time model introduced in \cite{bouchard_2015_arbit_nondo}.
Fix a time horizon \(N \in \mathbb{N}\), and let \(\mathbb{T} := \left\{ 0, 1, \dots, N \right\}\) be the time periods in this model.
Let \(\Omega_0 = \left\{ \omega_0 \right\}\) be a singleton and \(\Omega_1\) be a Polish space.
For each \(k \in \left\{ 1, \dots, N \right\}\), define \(\Omega_k := \Omega_0 \times \Omega_1^k\) as the \(k\)-fold Cartesian product.
For each \(k\), define \(\mathcal{G}_k := \mathcal{B}(\Omega_k)\) and let \(\mathcal{F}_k\) be its universal completion. 
In particular, we see that \(\mathcal{G}_0\) is trivial and denote \(\Omega := \Omega_N\) and \(\mathcal{F} := \mathcal{F}_N\). \footnote{We sometimes need \(\mathcal{G}_{k}\) via the embedding \(\mathcal{G}_{k} := \left\{ B \times \Omega_{N - k + 1}: B \in \mathcal{B} (\Omega_{k}) \right\}\) and \(\mathcal{F}_{k}\) the universal completion of \(\mathcal{G}_{k}\); for instance, in the conditional expectation \(\mathbb{E}^{Q} \left[ \Delta S_{k} \mid \mathcal{F} _{k - 1} \right]\) in \Cref{eq:def_mg}. Let us also emphasize that unless otherwise indicated, at any time \(k\), any mapping that takes in \(\Omega\) takes indeed the truncated \(\Omega_{k}\).}
We shall often see \(\left( \Omega_k, \mathcal{F}_k \right)\) as a subspace of \(\left( \Omega, \mathcal{F} \right)\), and \(\sigma\)-algebra \(\mathcal{F}_k\) as a sub-\(\sigma\)-algebra of \(\mathcal{F}_N\),
and hence obtain a filtration \(\mathbb{F} = \left( \mathcal{F}_k \right)_{0 \le k \le N}\) on \(\Omega\). 

Consider a market with \(d \in \mathbb{N}\) financial assets that can be traded dynamically without transaction costs. We model the dynamically traded assets by an \(\mathbb{R}^{d}\)-valued process \(S = \left( S_{t} \right)_{t \in \mathbb{T}}\) such that \(S_{t}\) is \(\mathcal{G}_t\)-measurable for \(t \in \mathbb{T}\). For an \(\mathbb{F}\)-predictable, \(\mathbb{R}^{d}\)-valued process \(H\), the terminal wealth of the hedging portfolio is given by
\[
\left(H \circ S \right)_{N} = \sum_{j = 1}^{d} \sum_{k = 1}^{N} H_{k}^{j} \Delta S_{k}^{j},
\]
where \(\Delta S_{k}^{j} = S_{k}^{j} - S_{k - 1}^{j}\).
Let \(\mathcal{A}\) be the space of actions at each time and let it be analytic and second-countable. 
We introduce \(\mathcal{C} := \mathcal{A}^{N+1}\) to be the collection of all possible plans, 
equipped with the Borel \(\sigma\)-algebra \(\mathcal{B}(\mathcal{C})\).

\begin{Example}
\label{exmp:swing}
Swing options in the natural gas market. When an industrial gas consumer wishes to hedge their future input costs, and she does not precisely know the consumption volume, but have an estimated range based on past usage, she could purchase a month-long swing option to offset the \textquotesingle volumetric risk\textquotesingle{} exposure,
The option will give her the right to buy up to \(10,000 \) MMBtu (one million British thermal unit) at a price \(p = \) \textsterling 5 in each of the four weeks, and the total volume must be between \(10,000 \) and \(30,000 \) MMBtu over the month.
In such a case, we could make our action space \(\mathcal{A} = \left[ 0, 10000 \right] \), the collection of all possible plans \(\mathcal{C} = \mathcal{A}^4 \) with the constraints that \(\sum_{i = 1}^{4} c_i \in \left[ 10000, 30000 \right] \).
\end{Example}

Let \(\tilde{\chi} : \Omega \times \mathcal{C} \times \mathbb{T} \rightarrow \mathcal{A}\) represent the feasible action sequence, and it is worth noticing that it is not only dependent on \(\omega\), but on previous actions as well. Then we have 
\begin{align}
\label{eq:def_d}
    \tilde{\mathcal{D}} &:= \Bigl\{ \tilde{\chi}: \Omega \times \mathcal{C} \times \mathbb{T} \to \mathcal{A} \mid \tilde{\chi}(\cdot, \cdot, k) ~\text{is}~ \mathcal{F}_k \otimes \mathcal{F}_{k}^{c} \text{-measurable}~ \forall k \in \left\{ 0, \dots, N \right\} ~\text{and}~ \nonumber \\[6pt]
&\phantom{:= \chi: \Omega \times \mathbb{T} \to \mathcal{A} | \chi(\cdot, k) ~\text{is}~ \mathcal{G}_k \text{-measurable}~ \forall k } \tilde{\chi}(\omega, c, \cdot) \in \mathcal{C}, \forall \omega \in \Omega, c \in \mathcal{C} \Bigr\}
\end{align}
representing the collection feasible action sequences, where \(\mathcal{F}_{k}^{c} := \sigma \left( \left( c_j \right)_{j \le k} \right)\) with \(c \in \mathcal{C}\), is the \(\sigma\)-algebra generated by the actions observed up to time \(k\).
We further define \(\chi : \Omega \times \mathbb{T} \rightarrow \mathcal{A}\) via  \(\chi (\cdot, 0) = \tilde{\chi} (\cdot, 0), \chi (\cdot, 1) = \tilde{\chi} (\cdot, \chi_{0}), \dots, \chi (\cdot, N) = \tilde{\chi} \bigl( \cdot, (\chi_{i})_{0 \le i \le N - 1} \bigr)\). Hence, we have

\[
\mathcal{D} := \left\{ \chi: \Omega \times \mathbb{T} \rightarrow \mathcal{A} \mid \chi (\cdot, k) ~\text{is}~ \mathcal{F}_{k} \text{-measurable}  \right\}.
\]

For a given \(k \in \left\{ 0, \dots, N - 1 \right\}\) and \(\omega \in \Omega_k\), we have a non-empty \textbf{convex} set \(\mathcal{P}_{k, k +1} (\omega) \subseteq \mathfrak{P}(\Omega_1)\) of probability measures, 
representing the set of all possible models for the \(\left( k + 1 \right)\)-th period, given the state \(\omega\) at time \(k\).
We assume that for each \(k \in \left\{ 0, \dots, N \right\}\),
\begin{equation}
\label{eq:graph_pk}
    \text{graph}(\mathcal{P}_{k, k + 1}) := \left\{ (\omega, \mathbb{P}): \omega \in \Omega_{k}, \mathbb{P} \in \mathcal{P}_{k, k + 1}(\omega) \right\} \subseteq \Omega_{k} \times \mathcal{P}(\Omega_{1}) ~\text{is analytic}.
\end{equation}
Given a universally measurable kernel \(\mathbb{P}_{k, k + 1}\) for each \(k \in \{ 0, 1, \dots, N - 1 \}\), define a probability measure \(\mathbb{P} = \mathbb{P}_{0, 1} \otimes \dots \otimes \mathbb{P}_{N - 1, N}\) on \(\Omega\) by 
\begin{equation}
\mathbb{P}(A) := \int _{\Omega_{1}} \dots \int _{\Omega_{1}} \mathbf{1}_{A} (\omega_{1}, \dots, \omega_{N}) \mathbb{P}_{N - 1, N}(\omega_{1}, \dots, \omega_{N - 1}; d\omega_{N}) \dots \mathbb{P}_{0, 1}(d\omega_{1}), \quad A \subseteq \Omega.
\end{equation}
We can then introduce the set \(\mathcal{P} \subseteq \mathfrak{P}(\Omega)\) of possible models for the multi-period market up to time \(N\) by 
\begin{equation}\label{eq:def_curlyp}
    \mathcal{P} := \left\{ \mathbb{P}_{0, 1} \otimes \mathbb{P}_{1, 2} \otimes \dots \otimes \mathbb{P}_{N - 1, N}: \mathbb{P}_{k, k + 1}(\cdot) \in \mathcal{P}_{k, k + 1}(\cdot), k = 0, 1, \dots, N - 1 \right\}.
\end{equation}
The set \(\mathcal{P}_{k, k + 1} (\omega)\) is non-empty, then by Jankov-von Neumann Theorem as in \cite[][Proposition 7.49, page 182]{bertsekas_2007_stoch_discr}, condition \Cref{eq:graph_pk} ensures that \(\mathcal{P}_{k, k + 1 }\) always has a universally measurable selector \(\mathbb{P}_{k, k + 1}: \Omega_{k} \to \mathfrak{P}(\Omega_{1})\) such that \(\mathbb{P}_{k, k + 1}(\omega) \in \mathcal{P}_{k, k + 1}(\omega)\) for all \(\omega \in \Omega_{k}\). 
Hence, the set \(\mathcal{P}\) defined in \Cref{eq:def_curlyp} in nonempty.

We introduce the set of one-step martingale measures dominated by elements of \(\mathcal{P}_{k, k + 1}(\omega)\)
\begin{equation}\label{eq:mk_one}
    \mathcal{M}_{k, k + 1}(\omega) := \left\{ \mathbb{Q} \in \mathfrak{P}(\Omega_{1}): \mathbb{Q} \lll \mathcal{P}_{k, k + 1}(\omega) ~\text{and}~ \mathbb{E}^{\delta_{\omega} \otimes_{k} \mathbb{Q}} [\Delta S_{k + 1}] = 0 \right\},
\end{equation}
where \(\Delta S_{k + 1} = S_{k + 1} - S_{k}\), and \(\delta_{\omega} \otimes_{k} \mathbb{Q} := \delta(\omega_{0}, \dots, \omega_{k}) \otimes \mathbb{Q}\) is a Borel probability measure on \(\Omega_{k + 1} := \Omega_{k} \times \Omega_{1}\). 
Furthermore,
\begin{equation}
\label{eq:mk}
    \mathcal{M}_{k}(\omega) := \left\{ \delta_{\omega} \otimes_{k} \mathbb{Q}_{k, k + 1} \otimes \dots \otimes \mathbb{Q}_{N - 1, N}: \mathbb{Q}_{i, i + 1}(\cdot) \in \mathcal{M}_{i, i + 1}(\cdot), i = k, \dots N - 1 \right\}. 
\end{equation}
The set of martingale measures for a process \(S\) on \(\Omega\) is given by 
\begin{equation} \label{eq:def_mg}
    \mathcal{M} = \left\{ \mathbb{Q} \in \mathfrak{P}(\Omega): \mathbb{Q} \lll \mathcal{P} ~\text{and}~ \mathbb{E}^{\mathbb{Q}} [\Delta S_{k} \mid \mathcal{F}_{k - 1}] = 0, \forall k = 1, \dots, N \right\},
\end{equation}
where \(\mathbb{Q} \lll \mathcal{P}\) denotes a stronger notion of absolute continuity, in the sense that
\[
\mathbb{Q} \lll \mathcal{P}, ~\text{if there exists}~ \mathbb{P} \in \mathcal{P} ~\text{such that}~ \mathbb{Q} \ll \mathbb{P}.
\]

Note that the family \(\big(\mathcal{M}_{k}(\omega) \big)_{k \leq N - 1, \omega \in \Omega}\) defined in \Cref{eq:mk} satisfies:
\begin{itemize}
\item \(\mathbb{Q} \big([\omega]_{\mathcal{G}_{k}}\big) = 1\) for all \(\mathbb{Q} \in \mathcal{M}_{k}(\omega)\), where \([\omega]_{\mathcal{G}_{k}}\) is the atom of \(\mathcal{G}_{k}\) containing \(\omega\); that is,
\begin{equation}\label{eq:om_atom}
    [\omega]_{\mathcal{G}_{k}} := \bigcap_{F \in \mathcal{G}_{k}: \omega \in F} F,
\end{equation}
is an indecomposable nonempty element of \(\mathcal{G}_{k}\), where \(\mathcal{G}_{k}\) is countably generated and hence \([\omega]_{\mathcal{G}_{k}} \in \mathcal{G}_{k}\); that is, the measures are concentrated on the paths of \(\Omega\) that coincides with \(\omega\) up until time \(k\);
\item For every \(\mathbb{Q} \in \mathcal{M}\) and every family of regular conditional probabilities \((\mathbb{Q}_{\omega})_{\omega \in \Omega}\) of \(\mathbb{Q}\) w.r.t. \(\mathcal{G}_{k}\), one has \(\mathbb{Q}_{\omega }\in \mathcal{M}_{k}(\omega)\), for \(\mathbb{Q}\)-a.e. \(\omega\).
\end{itemize}

\paragraph{The superhedging problem}
\label{sec:orgd10e3ef}
Let \(\Phi: \Omega \times \mathcal{C} \to \overline{\mathbb{R}}\) be the payoff function, and define the set of dynamic trading strategies
\begin{align}
\label{eq:strat}
    \mathcal{H} &:= \Bigl\{ H: \Omega \times \mathcal{C} \times \mathbb{T} \to \mathbb{R}^d \mid H(\cdot, \cdot, k + 1) =: H_{k + 1}(\cdot, \cdot) ~\text{is}~  \nonumber \\[6pt]
&\phantom{\Bigl\{ H: \Omega \times \mathcal{C} \times \mathbb{T} \to \mathbb{R}^d | H(\cdot, c, k + } \mathcal{F}_k \otimes \mathcal{F}_k^c \text{-measurable}, ~k = 0, \dots, N - 1 \Bigr\},
\end{align}
The superhedging price of the contract \(\Phi\) is given by 
\begin{equation}
\label{eq:sprice}
    \pi^{G}(\Phi) := \inf\left\{ x: \exists H \in \mathcal{H}, ~\text{s.t.,}~ x + \left( H(\cdot, c) \circ S \right)_{N} \geq \Phi(\cdot, c) ~\mathcal{P}\text{-q.s.,} \forall c \in \mathcal{C} \right\}.
\end{equation}

Recall that for a given \(\chi \in \mathcal{D}\) defined in \Cref{eq:def_d}, we have \(\chi(\omega, c, \cdot) \in \mathcal{C}\) for all \(\omega \in \Omega\), \(c \in \mathcal{C}\). 
Hence, we define \(\Phi_{\chi}: \Omega \to \overline{\mathbb{R}}\) by \(\Phi_{\chi}(\omega) = \Phi\left(\omega, \chi(\omega, \cdot, \cdot)\right)\). 
The model price is defined by
\begin{equation}
\label{eq:model_price}
    \sup_{\mathbb{Q} \in \mathcal{M}} \sup_{\chi \in \mathcal{D}} \mathbb{E}^{\mathbb{Q}} \left[\Phi_{\chi} \right].
\end{equation}

\begin{Remark}
Our setup allows restrictions on the available actions at each time. 
Moreover, such restrictions can also be dependent on \(\omega\), and whenever these restrictions are violated, the payoff function is set to \(-\infty\);
that is, there could be some subset of \(\Omega \times \mathcal{C}\) in the form \(\left\{ (\omega, c): \Phi(\omega, c) = -\infty \right\}\).

Here is an example of what we meant. 
For a three-period American option, our action space \(\mathcal{A} = \{ 0, 1 \}\), where \(0\) denotes \textquotesingle to continue\textquotesingle{} and \(1\) denotes \textquotesingle to stop\textquotesingle{}.
Since an American option can only be exercised once at any of the three periods, or not be exercised at all, 
any action sequence with more than one \textquotesingle 1\textquotesingle{} would be irrelevant, for example, \(\{ 0, 1, 1 \}\) or \(\{ 1, 1, 1 \}\), and its corresponding payoff would be set to \(-\infty\).

For another example, consider the situation where two American options are sold, with the restriction that they cannot be exercised at the same time. 
Denote by \(\tau_{1}\) and \(\tau_{2}\) their corresponding stopping times,  and we have that \(\Phi(\omega, \tau_{1}, \tau_{2}) = -\infty\) for \(\tau_{1} = \tau_{2}\).
\label{rm:infty}
\end{Remark}

\paragraph{The enlarged space}
\label{sec:orgf845312}
Consider the space \(\overline{\Omega} := \Omega \times \mathcal{C}\) with \(\bar{\omega} = (\omega, c) \in \overline{\Omega}\), where \(\omega \in \Omega\) and \(c \in \mathcal{C}\).
For each \(k \in \mathbb{T}\), define a restricted enlarged space 
\begin{equation}
\label{eq:res_enlarged}
    \overline{\Omega}_{k} := \Omega_{k} \times \mathcal{C}_{k - 1} \quad\text{and}\quad \overline{\Omega}_{k}^{+} := \Omega_{k} \times \mathcal{C}_{k},
\end{equation}
where \(\mathcal{C}_{k - 1} = \mathcal{A}^{k - 1}\), and \(\mathcal{C}_{-1}\) is defined to be the null action space.
We introduce the enlarged filtration \(\overline{\mathbb{G}} = (\overline{\mathcal{G}}_{k})_{0 \leq k \leq N}\), with
\[
\overline{\mathcal{G}}_{k} = \left\{ A \times \Omega_{N - k + 1} \times B \times \mathcal{A}^{N - k + 1}: A \in \mathcal{B}(\Omega_{k}), B \in \mathcal{B}(\mathcal{C}_{k + 1}) \right\} \subseteq \mathcal{B}(\overline{\Omega}), \quad  0 \le k \le N,
\]
\[
\overline{\mathcal{G}}_{k}^{-} = \left\{ A \times \Omega_{N - k + 1} \times B \times \mathcal{A}^{N - k + 2}: A \in \mathcal{B}(\Omega_{k}), B \in \mathcal{B}(\mathcal{C}_{k}) \right\} \subseteq \mathcal{B}(\overline{\Omega}), \quad  1 \le k \le N,
\]
and the universally completed filtration \(\overline{\mathbb{F}} = (\overline{\mathcal{F}}_{k})_{0 \leq k \leq N}\) with \(\overline{\mathcal{F}}_{k}\) being the universal completion of \(\overline{\mathcal{G}}_{k}\).
We think of \(\left( \overline{\Omega}_k, \overline{\mathcal{F}}_k \right)\) as a subspace of \(\left( \overline{\Omega}, \overline{\mathcal{F}} \right)\), and \(\sigma\)-algebra \(\overline{\mathcal{F}}_k\) as a sub-\(\sigma\)-algebra of \(\overline{\mathcal{F}}_N\).
Furthermore, for \(\bar{\omega} = (\omega, c) \in \overline{\Omega}_{k}\), define
\begin{align}
   \overline{\mathcal{P}}_{k, k + 1} (\omega, c) &= \Bigl\{ \overline{\mathbb{P}} \in \mathfrak{P}(\Omega_1 \times \mathcal{A}): \exists \nu \in \mathfrak{P} ~\text{s.t.}~ \overline{\mathbb{P}}(B \times A) = \int_A \mathbb{P}_a (B) \nu(da), \forall B \in \mathcal{B}(\Omega_1),
   \nonumber \\[6pt]
   & \phantom{ \Bigl\{ \overline{\mathbb{P}} \in \mathfrak{P}(\Omega_1 \times \mathcal{A}): \exists \nu \in \mathfrak{P} ~\text{s.t.}~ } A \subseteq \mathcal{A}, ~\text{and}~ \mathbb{P}_a \in \mathcal{P}_{k, k + 1}(\omega), ~\text{for \( \nu \)-a.e.}~ a \in A \Bigr\},
\end{align}
and 
\begin{equation*}
\label{eq:p_bar}
    \overline{\mathcal{P}} = \left\{ \overline{\mathbb{P}}_{0, 1} \otimes \dots \otimes \overline{\mathbb{P}}_{N - 1, N} \otimes \nu_{N}: \overline{\mathbb{P}}_{k, k + 1}(\cdot) \in \overline{\mathcal{P}}_{k, k + 1}(\cdot), k = 0, 1, \dots, N - 1, \nu_{N} \in \mathfrak{P}(\mathcal{A}) \right\}.
\end{equation*}
Then, by Fubini's Theorem, for all \(A \subseteq \overline{\Omega}\)
\begin{align*}
    & \overline{\mathbb{P}}_{0, 1} \otimes \dots \otimes \overline{\mathbb{P}}_{N - 1, N} \otimes \nu_{N}(A) \\[6pt]
    = & \int_{\Omega_{1} \times \mathcal{A}} \dots \int_{\Omega_{1} \times \mathcal{A} } \int_{\mathcal{A}} \mathbf{1}_A(\omega_{1}, \dots, \omega_{N}, c_{0}, \dots, c_{N - 1}, c_{N}) \\[6pt]
    & \quad \quad \quad \nu_{N}(dc_{N}) \overline{\mathbb{P}}_{N - 1, N}(\omega_{1}, \dots, \omega_{N - 1}, c_{0}, \dots, c_{N - 2}; d\omega_{N}, dc_{N - 1}) \dots \overline{\mathbb{P}}_{0, 1}(d\omega_{1}, dc_{0}) \\[6pt]
    = & \underbrace{\int_{\mathcal{A}} \dots \int_{\mathcal{A}}}_\text{\(N + 1 \)} \underbrace{\int_{\Omega_{1}} \dots \int_{\Omega_{1}}}_\text{\(N \)} \mathbf{1}_A(\omega_{1}, \dots, \omega_{N}, c_{0}, \dots, c_{N - 1}) \mathbf{1}_A(c_{N}) \\[6pt]
    & \quad \quad \quad \quad \quad \quad ~~~\mathbb{P}_{N - 1, N; c}(\cdot~; d\omega_{N}) \dots \mathbb{P}_{0, 1; c}(d\omega_{1}) \nu_{N}(dc_{N}) \nu_{N - 1}(dc_{N - 1}) \dots \nu_{0}(dc_{0}) \\[6pt]
    = & \int_{\mathcal{C}} \mathbb{P}_{c}(A_{c}) \nu(dc),
\end{align*}
where \(A_c := \left\{ \omega: (\omega, c) \in A, c \in \mathcal{C} \right\}\).

Hence, we may rewrite \(\overline{\mathcal{P}}\) as
\begin{align}
\label{eq:o_mtg}
    \overline{\mathcal{P}} &= \Bigl\{ \overline{\mathbb{P}} \in \mathfrak{P}(\overline{\Omega}): \exists \nu \in \mathfrak{P}(\mathcal{C}) ~\text{s.t.}~ \overline{\mathbb{P}}(B \times C) = \int_{C} \mathbb{P}_c(B) \nu(dc), \forall B \in \mathcal{F}, C \subseteq \mathcal{C}
    \nonumber \\[6pt]
    &\phantom{\Bigl\{ \overline{\mathbb{P}} \in \mathfrak{P}(\overline{\Omega}): \exists \nu \in \mathfrak{P}(\mathcal{C}) ~\text{s.t.}~ \overline{\mathbb{P}}(B \times \mathcal{C}) = \int_{C} \mathbb{P}_c(B) \nu(d} \mathbb{P}_{c} \in \mathcal{P}, ~\text{for \(\nu \)-a.e.}~ c \in C \Bigr\}.
\end{align}

We extend naturally the definition of \(S\) from \(\Omega\) to \(\overline{\Omega}\) via \(S(\bar{\omega}) = S(\omega)\) for \(\bar{\omega} = (\omega, c) \in \overline{\Omega}\).
We extend the definitions of \(\mathcal{M}_{k, k + 1}(\omega)\) and \(\mathcal{M}_{k}(\omega)\) defined in \eqref{eq:mk_one} and \eqref{eq:mk} respectively to the space \(\overline{\Omega}\) as follows. 
For \(\bar{\omega} = (\omega, c) \in \overline{\Omega}_{k}\), denote 
\begin{equation*} 
    \overline{\mathcal{M}}_{k, k + 1}(\bar{\omega}) := \left\{ \overline{\mathbb{Q}} \in \mathfrak{P}(\Omega_{1} \times \mathcal{A}): \overline{\mathbb{Q}} \lll \overline{\mathcal{P}}_{k, k + 1}(\omega, c) ~\text{and}~ \mathbb{E}^{\delta_{\bar{\omega}} \otimes_{k} \overline{\mathbb{Q}}} [\Delta S_{k + 1}] = 0  \right\},
\end{equation*} 
and 
\begin{equation}
\label{eq:mk_bar}
    \overline{\mathcal{M}}_{k}(\bar{\omega}) := \left\{ \delta_{\bar{\omega}} \otimes_{k} \overline{\mathbb{Q}}_{k, k + 1} \otimes \dots \otimes \overline{\mathbb{Q}}_{N - 1, N}: \overline{\mathbb{Q}}_{i, i + 1}(\cdot) \in \overline{\mathcal{M}}_{i, i + 1}(\cdot), i = k, \dots, N - 1 \right\}.
\end{equation} 
Analogous to \(\mathcal{M}\), the set of martingale measures on the enlarged space is as follows.
\begin{equation}
\label{eq:def_mgbar}
    \overline{\mathcal{M}} = \left\{ \overline{\mathbb{Q}} \in \mathfrak{P}(\overline{\Omega}): \overline{\mathbb{Q}} \lll \overline{\mathcal{P}} ~\text{and}~ \mathbb{E}^{\overline{\mathbb{Q}}} [\Delta S_{k} \mid \overline{\mathcal{F}}_{k - 1}] = 0, \forall k \in \{1, \dots, N \} \right\},
\end{equation}
The set of dynamic trading strategies is defined as
\begin{equation}
\label{eq:o_strat}
    \overline{\mathcal{H}} := \left\{ \overline{H}: \overline{\Omega} \times \mathbb{T} \to \mathbb{R}^{d} \mid \overline{H}(\cdot, k + 1) =: \overline{H}_{k + 1} ~\text{is \(\overline{\mathcal{F}}_{k} \)-measurable} \right\}.
\end{equation}

Let \(\overline{\Upsilon}\) be the class of upper semianalytic functions \(f: \overline{\Omega} \to \overline{\mathbb{R}}\) that are bounded from above. 
The following proposition states that the superhedging price of the contract \(\Phi\) is the same as that of its European option equivalence in the enlarged space.
\begin{Proposition}
For any \(\Phi \in \overline{\Upsilon}\),
\[
    \pi^G (\Phi) = \overline{\pi}^E (\Phi) := \inf \left\{x: \exists\overline{H} \in \overline{\mathcal{H}} ~\text{s.t.}~ x + (\overline{H} \circ S)_N \ge \Phi ~\overline{\mathcal{P}}\text{-q.s.} \right\}
    \]
\label{prop:equiv_price}
\end{Proposition}

\begin{proof}
From \Cref{eq:strat,eq:o_strat}, note that the dynamic trading strategies for superhedging in \(\pi^{G}_{}\) and \(\overline{\pi}^{E}_{ }\) are the same.
Then for a set \(\Gamma \in \overline{\mathcal{F}}_{N}\), let \(\Gamma_{c} = \left\{ \omega: (\omega, c) \in \Gamma \right\}\) and assume that \(\overline{\mathbb{P}}(\Gamma) = 0\) for every \(\overline{\mathbb{P}} \in \overline{\mathcal{P}}\).
For any \(\mathbb{P} \in \mathcal{P}\) and \(c \in \mathcal{C}\), define \(\overline{\mathbb{P}} := \mathbb{P} \otimes \delta_{c} \in \overline{\mathcal{P}}\), thus \(\mathbb{P}(\Gamma_{c}) = 0\).
Therefore, for \(x\) and \(\overline{H}\) such that \(x + (\overline{H} \circ S)_{N} \geq \Phi\), \(\overline{\mathbb{P}}\)-a.s., 
we have that \(x + (H(\cdot, c) \circ S)_{N} \geq \Phi(\cdot, c)\), \(\mathbb{P}\)-a.s., for any \(c \in \mathcal{C}\) and \(\mathbb{P} \in \mathcal{P}\). 
Hence, we have \(\overline{\pi}^{E}_{}(\Phi) \geq \pi^{G}_{}(\Phi)\).

For the other direction, note that \(\overline{\mathbb{P}}(\Gamma) = \int_{\mathcal{C}} \mathbb{P}_{c}(\Gamma_{c}) d\nu(c)\) where \(\mathbb{P}_{c} \in \mathcal{P}\) by construction.
Hence, by the same arguments as above, it follows \(\overline{\pi}^{E}_{}(\Phi) \leq \pi^{G}_{}(\Phi)\). 
\end{proof}

\section{Pricing-hedging duality}
\label{sec:main1}
Our main result is the European pricing-hedging duality on the enlarged space, 
which is stated in \Cref{thm:dual_bar}. 
A dynamic programming principle result on the enlarged space is proved in \Cref{prop:dpp_bar}.

Recall the notion of no arbitrage introduced in \cite{bouchard_2015_arbit_nondo},
stating that \(NA \left(\mathcal{P} \right)\) holds if for all \(H \in \mathcal{H}\),
\[
\left(H \circ S \right)_{N } \ge 0 \quad \mathcal{P}\text{-q.s.} \implies \left(H \circ S \right)_{N } = 0 \quad \mathcal{P}\text{-q.s.}
\]
Analogously, we say that \(NA(\overline{\mathcal{P}})\) holds if for \(\overline{H} \in \overline{\mathcal{H}}\),
\begin{equation}
\label{eq:na_bar}
    \left(\overline{H} \circ S \right)_{N} \geq 0 \quad \overline{\mathcal{P}}\text{-q.s.} \implies \left(\overline{H} \circ S \right)_{N} = 0 \quad \overline{\mathcal{P}}\text{-q.s.}
\end{equation}

Recall the sets \(\mathcal{M}_{}\) and \(\overline{\mathcal{M}}_{}\) defined in \Cref{eq:def_mg,eq:def_mgbar}.
Bouchard and Nutz \cite{bouchard_2015_arbit_nondo} established that \(NA \left(\mathcal{P} \right)\) is equivalent to \(\mathcal{P}\) and \(\mathcal{M}_{}\) having the same polar sets.
We will obtain our version of such an equivalence on \(\overline{\Omega}\).
Then we follow the same arguments as in the proof of \Cref{prop:equiv_price} above, and obtain the following lemma.

\begin{Lemma}
\(NA \left(\mathcal{P} \right) \iff NA \left(\overline{\mathcal{P}} \right) \iff \overline{\mathcal{P}}\) and \(\overline{\mathcal{M}}_{}\) have the same polar sets.
\end{Lemma}

Under the no-arbitrage condition \Cref{eq:na_bar}, we have the following superhedging duality.
\begin{Theorem}
Let \(NA \left(\overline{\mathcal{P}} \right)\) hold, and let \(\Phi: \overline{\Omega} \to \overline{\mathbb{R}}\) be upper semianalytic. One has
\begin{equation}
\label{eq:bar_price}
    \overline{\pi}^{E}_{}(\Phi) = \sup_{\mathbb{Q} \in \mathcal{M}} \sup_{\chi \in \mathcal{D}} \mathbb{E}^{\mathbb{Q}} \left[ \Phi_{\chi} \right] = \sup_{\overline{\mathbb{Q}} \in \overline{\mathcal{M}}} \mathbb{E}^{\overline{\mathbb{Q}}} \left[\Phi \right].
\end{equation}
Furthermore, there exists \(\overline{H} \in \overline{\mathcal{H}}\) such that
\[
        \overline{\pi}^E (\Phi) + \left(\overline{H} \circ S \right)_N \ge \Phi, \quad \overline{\mathcal{P}}\text{-q.s.}
    \]
\label{thm:dual_bar}
\end{Theorem}

\begin{Remark}
A probability measure \(\overline{\mathbb{Q}} \in \overline{\mathcal{M}}\) can be seen as a market model, together with a sequence of actions \(c \in \mathcal{C}\). Indeed, the set \(\overline{\mathcal{M}}\) is larger than the set of all push-forward measures induced by \(\omega \mapsto (\omega, \chi(\omega, \cdot))\) and \(\mathbb{Q} \in \mathcal{M}\) for all \(\chi \in \mathcal{D}\). In addition to all feasible action sequences in \(\mathcal{D}\), it also takes into consideration all randomized plans in \(\mathcal{C}\). Hence we have \emph{a weak formulation} of the problem \Cref{eq:model_price}. 
\end{Remark}

\begin{Remark}
Suppose additionally there are \(e \in \mathbb{N}\) European options available for static trading in the market. We use a \(\mathbb{R}^{e}\)-valued, \(\mathcal{G}_{N}\)-measurable random vector \(\textsl{g} = \left( \textsl{g}^{1}, \dots, \textsl{g}^{e} \right)\) to represent the payoffs of these statically traded European options. Without loss of generality we assume the prices of these European options are \(0\). As shown in \cite{bayraktar_2013_hedgi_uncer}, when statically traded options are included in the hedging portfolio, it is possible that
\[
        \pi^G_{\textsl{g}} (\Phi) = \overline{\pi}^E_{\textsl{g}} (\Phi) \ne \sup_{\mathbb{Q} \in \mathcal{M}_{\textsl{g}}} \sup_{\chi \in \mathcal{D}} \mathbb{E}^{\mathbb{Q}} \left[\Phi_{\chi} \right],
    \]
where
    \begin{align*}
\pi_{\textsl{g}}^{G} (\Phi) &:= \inf \left\{x: \exists (H, h) \in \mathcal{H} \times \mathbb{R}^{e}, ~\text{s.t.,}~ x + \left( H(\cdot, c) \circ S \right)_{N} + h\textsl{g} \ge \Phi(\cdot, c) ~\mathcal{P}\text{-q.s.,} \forall c \in \mathcal{C} \right\},\\[6pt]
\overline{\pi}_{\textsl{g}}^E (\Phi) &:= \inf \left\{x: \exists (\overline{H}, h) \in \overline{\mathcal{H}} \times \mathbb{R}^{e} ~\text{s.t.,}~ x + \left( \overline{H} \circ S \right)_{N} + h\textsl{g} \ge \Phi ~\overline{\mathcal{P}}\text{q.s.} \right\},
\end{align*}
are the superhedging prices in the original and the enlarged space respectively, and
\[
\mathcal{M}_{\textsl{g}} = \left\{ \mathbb{Q} \in \mathcal{M}: \mathbb{E}^{\mathbb{Q}} \left[ \textsl{g}^{i} \right] = 0, i = 1, \dots, e \right\}.
\]
That is, there could be a duality gap between the superhedging price and model price in the original space.

To close the duality gap, one solution is to establish the duality on the enlarged probability space, i.e.,
\begin{equation}
\label{eq:rm_dual}
\overline{\pi}^{E} (\Phi) = \sup_{\overline{\mathbb{Q}} \in \overline{\mathcal{M}}_{\textsl{g}}} \mathbb{E}^{\overline{\mathbb{Q}}} \left[ \Phi \right],
\end{equation}
where
\[
\overline{\mathcal{M}}_{\textsl{g}} = \left\{ \overline{\mathbb{Q}} \in \overline{\mathcal{M}}: \mathbb{E}^{\overline{\mathbb{Q}}} \left[ \textsl{g}^{i} \right] = 0, i = 1, \dots, e \right\}
\]
with \(\textsl{g}\) being extended to the enlarged space in an obvious way. Another solution relies on a dynamic extension of the original market such that the European options can also be dynamically traded as that in \cite[][Section 2.3, Page 871]{aksamit_2019_robus_prici}. Then we would obtain a duality result of the form:
\begin{equation}
\label{eq:rm_dual_g}
\pi_{\textsl{g}}^G(\Phi) = \widehat{\pi}^G(\Phi) = \sup_{\widehat{\mathbb{Q}} \in \widehat{\mathcal{M}}} \sup_{\widehat{\chi} \in \widehat{\mathcal{D}}} \mathbb{E}^{\widehat{\mathbb{Q}}} \left[\Phi_{\widehat{\chi}} \right] = \sup_{\widehat{\mathbb{Q}} \in \overline{\widehat{\mathcal{M}}}} \mathbb{E}^{\widehat{\mathbb{Q}}} \left[ \Phi \right],
\end{equation}
where  \(\left( \widehat{\Omega}, \widehat{\mathbb{F}}, \widehat{\mathcal{F}}, \widehat{\mathcal{P}}, Y \right)\) is a dynamic extension of \(\left( \Omega, \mathbb{F}, \mathcal{F}, \mathcal{P}, S, \textsl{g} \right)\), with \(Y\) being the price process of European options \(\textsl{g}\), and the  \(\widehat{\phantom{a}}\) and \(\overline{\phantom{a}}\) represent the space enlargement by the price processes of European options and the action space, respectively. (See \cite[][Section 2.3]{aksamit_2019_robus_prici} for more details.)
The proof of \Cref{eq:rm_dual} and \Cref{eq:rm_dual_g} uses an inductive argument on the number of European options. We omit the proof details here since such argument would be similar to that in e.g., \cite[][Section 5]{bouchard_2015_arbit_nondo} or \cite[][Section 6.2]{aksamit_2019_robus_prici}. 
\label{rm:dym}
\end{Remark}

\subsection{Proof of \Cref{thm:dual_bar}}
\label{sec:org7d2b722}
To prove \Cref{thm:dual_bar}, we first introduce the following set of measures
\begin{equation*}
    \overline{\mathcal{M}}^{loc} := \left\{ \overline{\mathbb{Q}}: \overline{\mathbb{Q}} \lll \overline{\mathcal{P}} ~\text{and}~ S ~\text{is an}~ \left(\overline{\mathbb{F}}, \overline{\mathbb{Q}} \right)\text{-local martingale} \right\}.
\end{equation*}

By similar arguments as those in \cite[][Lemma A.2 and Lemma A.1]{bouchard_2015_arbit_nondo} and \cite[][Lemma 6.3]{aksamit_2019_robus_prici}, we have the following two lemmas.
\begin{Lemma}
Let \(\Phi\) be upper semianalytic and \(\overline{\mathbb{Q}} \in \overline{\mathcal{M}}^{loc}\). 
Then, for any \(x \in \mathbb{R}\) and \(\overline{H} \in \overline{\mathcal{H}}\) such that 
\(x + \left(\overline{H} \circ S \right)_{N}(\omega, c) \geq \Phi(\omega, c)\), \(\overline{\mathbb{Q}}\)-a.s., one has \(\mathbb{E}^{\overline{\mathbb{Q}}} \left[\Phi \right] \leq x\).
\label{lm:weak_dual}
\end{Lemma}

\Cref{lm:weak_dual} gives us the weak duality for all upper semianalytic \(\Phi\)
\begin{equation}
\label{eq:weak_dual}
    \sup_{\overline{\mathbb{Q}} \in \overline{\mathcal{M}}} \mathbb{E}^{\overline{\mathbb{Q}}} [\Phi] \leq \sup_{\overline{\mathbb{Q}} \in \overline{\mathcal{M}}^{loc}} \mathbb{E}^{\overline{\mathbb{Q}}} [\Phi] \leq \overline{\pi}^{E}(\Phi).
\end{equation}

\begin{Lemma}
Let \(\Phi\) be upper semianalytic, let \(\overline{\mathbb{Q}} \in \overline{\mathcal{M}}^{loc}\) 
and \(\phi: \overline{\Omega} \to [1, \infty)\) be a random variable such that \(\lvert \Phi(\omega, c)\rvert \leq \phi(\omega)\) for all \(\bar{\omega} = (\omega, c) \in \overline{\Omega}\). 
Then, we have that \(\overline{\mathcal{M}}^{\phi, \overline{\mathbb{Q}}} \neq \emptyset\) and that
\[
        \mathbb{E}^{\overline{\mathbb{Q}}} \left[\Phi \right] \le \sup_{\overline{\mathbb{Q}}^{\prime} \in \overline{\mathcal{M}}^{\phi, \overline{\mathbb{Q}}}} \mathbb{E}^{\overline{\mathbb{Q}}'} \left[\Phi \right],
    \]
where
\[
        \overline{\mathcal{M}}^{\phi, \overline{\mathbb{Q}}} := \left\{ \overline{\mathbb{Q}}^{\prime} \sim \overline{\mathbb{Q}}: \mathbb{E}^{\overline{\mathbb{Q}}^{\prime}} \left[\Phi \right] < \infty, ~\text{and}~ S ~\text{is an}~ \left(\overline{\mathbb{F}}, \overline{\mathbb{Q}}^{\prime} \right)\text{-martingale}  \right\}.
    \]   
\label{lm:mtg_intg}
\end{Lemma}

For \(1 \leq i \leq j \leq N\), define the map from \(\Omega_{j}\) to \(\Omega_{i}\) (respectively, \(\overline{\Omega}_{j}\) to \(\overline{\Omega}_{i}\)) by
\[
\left[ \omega \right]_{i} := \left( \omega_{1 } , \dots, \omega_{i }\right), \quad  \forall \omega \in \Omega_{j }
\]
and
\[
\left[ \bar{\omega} \right]_{i} := \left( [\omega]_{i } , [c]_{i - 1}\right), \quad  [c]_{i} = (c_{l })_{l \le i}, \quad  \forall \bar{\omega} = (\omega, c) \in \overline{\Omega}_{j}.
\]
Moreover, the map from \(\overline{\Omega}_{j}\) to \(\overline{\Omega}_{i}^{+}\) is defined by 
\[
\left[ \bar{\omega} \right]^{+}_{i} := \left( [\omega]_{i } , [c]_{i }\right), \quad  \forall \bar{\omega} = (\omega, c) \in \overline{\Omega}_{j},
\]
Note that \(\overline{\mathcal{G}}_{k}^{-}\) is the smallest \(\sigma\)-algebra on \(\overline{\Omega}\) generated by \([\cdot]_{k}: \overline{\Omega} \to \overline{\Omega}_{k}\); 
a \(\overline{\mathcal{G}}_{k}^{-}\)-measurable random variable \(f\) on \(\overline{\Omega}\) can be identified as a Borel measurable function on \(\overline{\Omega}_{k}\).

Denote \(\mathcal{E}(\xi) := \sup_{\mathbb{Q} \in \mathcal{M}} \mathbb{E}^{\mathbb{Q}} [\xi]\) for \(\xi \in \Upsilon\), where \(\Upsilon\) is the collection of upper semianalytic functions \(f: \Omega \to \overline{\mathbb{R}}\).
Suppose there is a family of operators \(\mathcal{E}_{k}: \Upsilon \to \Upsilon\) for \(k \leq N\), defined as
\begin{align*}
    \mathcal{E}_{k}(\xi)(\omega) &= \sup_{\mathbb{Q} \in \mathcal{M}_{k}(\omega)} \mathbb{E}^{\mathbb{Q}} \left[\xi \right], \quad k \leq N - 1, \\[6pt]
    \mathcal{E}_{N}(\xi)(\omega) &= \xi(\omega),
\end{align*}
such that \(\mathcal{E}_{k}(\xi)\) is \(\mathcal{F}_{k}\)-measurable for all \(\xi \in \Upsilon\) and \(\mathcal{E}_{k}(-\infty) = -\infty\). 
By \((4.12)\) of \cite{bouchard_2015_arbit_nondo}, we have that \((\mathcal{E}_{k})\) provides a dynamic programming representation of \(\mathcal{E}\);
that is,
\begin{equation}
\label{eq:dpp}
    \mathcal{E}(\xi) = \mathcal{E}^{0}(\xi) ~~\forall \xi \in \Upsilon, ~\text{where}~ \mathcal{E}^{k}(\xi) := \mathcal{E}_{k} \circ \dots \circ \mathcal{E}_{N}(\xi), \quad 0 \leq k \leq N.
\end{equation}
Observe that
\begin{equation}
\label{eq:eqv_e}
    \mathcal{E}_{k} \circ \dots \circ \mathcal{E}_{N - 1}(\xi)(\omega) = \mathcal{E}_{k, k + 1} \circ \dots \circ \mathcal{E}_{N - 1, N}(\xi)(\omega), \quad \xi \in \Upsilon,
\end{equation} 
where \(\mathcal{E}_{k, k + 1}(\xi)(\omega) = \sup_{\mathcal{M}_{k, k + 1}(\omega)} \mathbb{E}^{\mathbb{Q}} \left[\xi \right]\).

We extend the family of operators \((\mathcal{E}_{k})\) to \((\overline{\mathcal{E}}_{k})\), in the following form.
For \(f: \overline{\Omega}_{k + 1} \to \mathbb{R}\), and \(\bar{\omega} = (\omega, c) \in \overline{\Omega}_{k}\),
\begin{equation}
\label{eq:e_bar}
    \overline{\mathcal{E}}_{k}(f)(\omega, c) = \sup_{\mathbb{Q} \in \mathcal{M}_{k}(\omega)} \sup_{c^{\prime} \in \mathcal{C}_{N - k + 1}} 
    \mathbb{E}^{\mathbb{Q}} \left[ f(\cdot, c, c^{\prime}) \right], \quad 0 \leq k \leq N - 1.
\end{equation}
In particular,
\begin{equation}
\label{eq:e}
    \overline{\mathcal{E}}_{N}(\Phi)(\omega, c) = \sup_{c' \in \mathcal{C}_{1}} \Phi(\omega, c, c').
\end{equation}
Also observe that
\begin{equation}
\label{eq:eqv_ebar}
    \overline{\mathcal{E}}_{k} \circ \dots \circ \overline{\mathcal{E}}_{N}(\Phi)(\omega, c) = \overline{\mathcal{E}}_{k, k + 1} \circ \dots \circ \overline{\mathcal{E}}_{N - 1, N} \circ \overline{\mathcal{E}}_{N}(\Phi)(\omega, c), \quad \Phi \in \overline{\Upsilon},
\end{equation}
where \(\overline{\mathcal{E}}_{k, k + 1}(f)(\omega, c) = \sup_{\mathbb{Q} \in \mathcal{M}_{k, k + 1}(\omega)} \sup_{c' \in \mathcal{C}_{N - k + 1}} \mathbb{E}^{\mathbb{Q}} \left[f(\omega, \cdot, c, c')\right]\), for \(f: \overline{\Omega}_{k + 1} \to \overline{\mathbb{R}}\). 

For \(\xi \in \overline{\Upsilon}\), let
\begin{equation}
\label{eq:e_k}
  \overline{\mathcal{E}}^{k}(\xi) := \overline{\mathcal{E}}_{k} \circ \dots \circ \overline{\mathcal{E}}_{N}(\xi), \quad 0 \leq k \leq N.
\end{equation}

\begin{Proposition}
Let \(\Phi: \overline{\Omega} \to \mathbb{R}\) be upper semianalytic and bounded from above, then 
    \begin{equation}
\label{eq:dpp_bar}
        \sup_{\overline{\mathbb{Q}} \in \overline{\mathcal{M}}} \mathbb{E}^{\overline{\mathbb{Q}}} [\Phi] = \overline{\mathcal{E}}(\Phi) = \overline{\mathcal{E}}^0(\Phi) = \overline{\mathcal{E}}_{0} \circ \dots \circ \overline{\mathcal{E}}_{N}(\Phi).
    \end{equation}
\label{prop:dpp_bar}
\end{Proposition}

To prove \Cref{prop:dpp_bar}, we first prove the following two lemmas to establish a connection between the families of operators \((\mathcal{E}_{k})\) and \((\overline{\mathcal{E}}_{k})\), hence the \textquotesingle \(\geq\)\textquotesingle{} side of the equality.

\begin{Lemma}
For any \(\varepsilon > 0\) and \(\varphi \in \overline{\Upsilon}\), there exists a \(\chi^{\varepsilon} \in \mathcal{D}\) such that
\[
        \overline{\mathcal{E}}^0(\varphi) \le \mathcal{E}^0(\varphi_{\chi^{\varepsilon}}) + \varepsilon.
    \]
\label{lm:ebar_eps}
\end{Lemma}

\begin{proof}
We adapt the proof of \cite[][Theorem 2.3]{nutz_2013_const_sublin}. Let \(\mathfrak{X} = \mathfrak{P} \times \Omega_{k} \times \mathcal{C}_{k - 1} \times \mathcal{A}\) and consider the mapping \(K : \mathfrak{X} \rightarrow \mathfrak{P} (\overline{\Omega}_{k + 1})\) defined by
\[
K(A; \mathbb{P}, \omega, c, c') = \mathbb{E}^{\mathbb{P}} \left[ \mathbf{1}_{A} (\omega, \cdot, c, c') \right], \quad A \in \mathcal{\mathcal{G}}_{k + 1}.
\]

Then, we consider \(\mathbb{P}\) as a measure on \(\mathcal{F}_{k + 1}\) and let us show that \(K\) is a Borel kernel; that is,
\[
K : \mathfrak{X} \rightarrow \mathfrak{P} (\overline{\Omega}_{k + 1}) \quad \text{is Borel measurable}.
\]
This is equivalent to saying that \((\mathbb{P}, \omega, c, c') \mapsto \mathbb{E}^{\mathbb{P}} \left[ f(\omega, \cdot, c, c') \right]\) is Borel measurable whenever \(f : \Omega_{k} \times \Omega_{1} \times \mathcal{C}_{k - 1} \times \mathcal{A} \rightarrow \mathbb{R}\) is bounded and Borel measurable. We now consider the set \(W\) of all bounded Borel functions \(g : \Omega_{k} \times \Omega_{1} \times \mathcal{C}_{k - 1} \times \mathcal{A} \rightarrow \mathbb{R}\) such that
\[
(\mathbb{P}, \omega, c, c') \mapsto \mathbb{E}^{\mathbb{P}} \left[ g(\omega, \cdot, c, c') \right] \quad \text{is Borel measurable for any}~ c \in \mathcal{C}_{k - 1},
\]
and show that it contains a monotone class. 

The set \(W\) is a linear space, contains all constants and if \(g_{n} \in W\) increases to a bounded function \(g\),  then the  above condition is satisfied as \(\left( \mathbb{P}, \omega, c, c' \right) \mapsto \mathbb{E}^{\mathbb{P}} \left[ g(\omega, \cdot, c, c') \right]\) is the pointwise limit of the Borel-measurable functions \(\left( \mathbb{P}, \omega, c, c' \right) \mapsto \mathbb{E}^{\mathbb{P}} \left[ g_{n} (\omega, \cdot, c, c') \right]\).
Moreover, \(W\) contains any bounded, uniformly continuous function \(g\). Indeed, if \(\rho\) is a modulus of continuity for \(g\) and \(\left( \mathbb{P}^{n}, \omega^{n}, c^{n}, c'^{n} \right) \rightarrow \left( \mathbb{P}, \omega, c, c' \right)\) in \(\mathfrak{X}\), then 
\begin{align*}
\Bigl\vert \mathbb{E}^{\mathbb{P}^{n}} [ g(\omega_{n}, \cdot, c_{n}, c'_{n}) ] - \mathbb{E}^{\mathbb{P}} [ g(\omega, \cdot, c, c') ] \Bigr\vert &\le \left\vert \mathbb{E}^{\mathbb{P}^{n}} \left[ g(\omega_{n}, \cdot, c, c') \right] - \mathbb{E}^{\mathbb{P}^{n}} \left[ g(\omega, \cdot, c, c') \right]\right\vert \\[6pt]
&\phantom{\le} + \left\vert \mathbb{E}^{\mathbb{P}^{n}} \left[ g(\omega, \cdot, c, c') \right] - \mathbb{E}^{\mathbb{P}} \left[ g(\omega, \cdot, c, c') \right]\right\vert \\[6pt]
&\phantom{\le} + \left\vert \mathbb{E}^{\mathbb{P}^{n}} \left[ g(\omega, \cdot, c, c'_{n}) \right] - \mathbb{E}^{\mathbb{P}^{n}} \left[ g(\omega, \cdot, c, c') \right]\right\vert \\[6pt]
&\phantom{\le} + \left\vert \mathbb{E}^{\mathbb{P}^{n}} \left[ g(\omega, \cdot, c_{n}, c') \right] - \mathbb{E}^{\mathbb{P}^{n}} \left[ g(\omega, \cdot, c, c') \right]\right\vert \\[6pt]
& \le \rho \left( \text{dist}(\omega^{n}, \omega) \right) + \rho \left( \text{dist}(c'^{n}, c') \right) + \rho \left( \text{dist}(c^{n}, c) \right) \\[6pt]
&\phantom{\le} + \left\vert \mathbb{E}^{\mathbb{P}^{n}} \left[ g(\omega, \cdot, c, c') \right] - \mathbb{E}^{\mathbb{P}} \left[ g(\omega, \cdot, c, c') \right]\right\vert \rightarrow 0,
\end{align*}
showing that \(\left( \mathbb{P}, \omega, c, c' \right) \mapsto \mathbb{E}^{\mathbb{P}} \left[ g(\omega, \cdot, c, c') \right]\) is continuous and thus Borel-measurable.
Since the uniformly continuous functions generate the Borel \(\sigma\)-algebra on \(\Omega_{k} \times \Omega_{1} \times \mathcal{C}_{k - 1} \times \mathcal{A}\), the monotone class theorem implies that \(W\) contains all bounded Borel-measurable functions.
Therefore, \(K\) is \(\sigma(W)\)-measurable; that is, \(K\) is a Borel kernel.
Then by \cite[][Proposition 7.48]{bertsekas_2007_stoch_discr}, which says that Borel kernels integrate upper semianalytic functions into upper semianalytic ones, the mapping
\[
\left( \mathbb{P}, \omega, c, c' \right) \mapsto \mathbb{E}^{\mathbb{P}} \left[ f(\omega, \cdot, c, c') \right] \equiv \int f(\omega, \omega', c, c') K(d\omega'; \mathbb{P}, \omega, c')
\]
is upper semianalytic.
After recalling that \(\mathcal{P}_{k, k + 1}\) has an analytic graph and \(\mathcal{A}\) is analytic, it follows from the Projection Theorem in form of \cite[][Proposition 7.47]{bertsekas_2007_stoch_discr}, that \(\overline{\mathcal{E}}_{k} (f) = \sup_{(\mathbb{P}, c)} \mathbb{E}^{\mathbb{P}} \left[ f(\omega, \cdot, c, \cdot) \right]\) is upper semianalytic (in \((\Omega_{1} \times \mathcal{A})\)). 
Then by the Jankov-von Neumann Theorem in the form of \cite[][Proposition 7.50, page 184]{bertsekas_2007_stoch_discr}, and Fubini's theorem
\begin{align*}
\left( \overline{\mathcal{E}}_{0, 1} \circ \dots \circ \overline{\mathcal{E}}_{N} \right) (\varphi) &= \overline{\mathcal{E}}_{0, 1} \left( \overline{\mathcal{E}}^{1} (\varphi) \right) \\[6pt]
& \le \mathbb{E}^{\mathbb{Q}^{\varepsilon}_{0}} \left[ \overline{\mathcal{E}}^{1} (\varphi) (\omega_{0}, \cdot, \cdot, c^{\varepsilon}_{0}) \right] + \varepsilon \\[6pt]
& \le \mathbb{E}^{\mathbb{Q}^{\varepsilon}_{0}} \left[ \mathbb{E}^{\mathbb{Q}^{\varepsilon}_{1}} \left[ \overline{\mathcal{E}}^{2} (\varphi) (\omega_{1}, \cdot, \cdot, c^{\varepsilon}_{1}) \right] \right] + 2\varepsilon \\[6pt]
&\phantom{abcdefg} \vdots \\[6pt]
& \le N\varepsilon + \mathbb{E}^{\mathbb{Q}^{\varepsilon}} \left[ \varphi(\cdot, \chi^{\varepsilon}) \right],
\end{align*}
which results in
\[
\overline{\mathcal{E}}^{0} (\varphi) \le N\varepsilon + \sup_{\mathbb{Q} \in \mathcal{M}} \mathbb{E}^{\mathbb{Q}} \left[ \varphi_{\chi^{\varepsilon}} \right],
\]
 with \(\chi^{\varepsilon} (\omega) := \left( c_{0}^{\varepsilon} (\omega_{0}), c_{1}^{\varepsilon} (\omega_{1}, c_{0}^{\varepsilon} (\omega_{0}))\dots, c_{N}^{\varepsilon} (\omega_{N}, c_{N - 1}^{\varepsilon} (\omega_{N - 1}, \dots) )\right)\) for any \(\omega \in \Omega\). \footnote{To help better picture \(\chi\), here is what it looks like in a three-period example, \(\chi(\omega) = \left( \chi_{0} (\omega_{0}), \underbrace{\chi_{1} (\omega_{1}, \chi_{0}))}_{c_{1}},  \underbrace{\chi_{2} (\omega_{2}, (\chi_{0}, \chi_{1})))}_{c_{2}} \right)\).}

To see that \(\chi^{\varepsilon}_{k}\) is indeed \(\mathcal{F}_{k}\)-measurable, we proceed inductively. By definition, \(c^{\varepsilon}_{0}\) is \(\mathcal{F}_{0}\)-measurable. Assuming that for any time \(k\), \(\chi^{\varepsilon}_{k}\) is \(\mathcal{F}_{k}\)-measurable, we look at
\[
c^{\varepsilon}_{k + 1} \bigl(\omega_{k + 1}, (\chi^{\epsilon}_{i})_{0 \le i \le k} \bigr),
\]
which is the universally (or analytically measurable) selector at time \(k + 1\); that is, \linebreak \(R := \left\{ (\omega_{k + 1}, c_k) \mid c^{\varepsilon}_{k + 1} (\omega_{k + 1}, c_{k}) \in B, B \in \mathcal{B}_{\mathcal{A}} \right\} \in \mathscr{A}_{\Omega_{k + 1} \times \mathcal{C}_{k}}\), the analytic \(\sigma\)-algebra. \footnote{We recall that for any Borel space \(X\), \(\mathcal{B}_{X} \subset \mathscr{A}_{X} \subset \mathscr{U}_{X}\), where \(\mathcal{B}_{X}\) is the Borel \(\sigma\)-algebra, \(\mathscr{U}_{X}\) the universal \(\sigma\)-algebra and \(\mathscr{A}_{X}\) the analytic \(\sigma\)-algebra. (See \cite[, page 171]{bertsekas_2007_stoch_discr}, for details.)} Hence, we have that  \(R_{\mid \Omega_{k + 1}} \in \mathcal{F}_{k +  1}\).
We claim that \(\chi^{\varepsilon}_{k}\) is also \(\mathcal{F}_{k + 1}\)-measurable. Indeed, via some embedding \((\chi^{\varepsilon}_{k})^{-1} (B) := \left\{ \omega \in \Omega_{k} \mid \chi^{\varepsilon}_{k} (\omega) \in B, B \in \mathcal{B}_{\mathcal{A}} \right\} \times \Omega_{1} \in \mathcal{F}_{k + 1}\).
Therefore, \(\chi^{\varepsilon}_{k + 1}\) is \(\mathcal{F}_{k + 1}\)-measurable.
\end{proof}

The next lemma follows directly from the proof of \Cref{lm:ebar_eps}. 
\begin{Lemma}
Let \(\varphi \in \overline{\Upsilon}\), then we have that \(\overline{\mathcal{E}}_{k} (f): \overline{\Omega}_{k} \rightarrow \overline{\mathbb{R}}\) is upper semianalytic, whenever \(f : \overline{\Omega}_{k + 1} \rightarrow \overline{\mathbb{R}}\) is upper semianalytic.
\label{lm:mrb_chi}
\end{Lemma}

Next, we show the \textquotesingle\(\leq\)\textquotesingle{} side of the equality in \Cref{eq:dpp_bar}. Recall the family of operators \(\left(\overline{\mathcal{E}}_{k}\right)\) defined in \Cref{eq:e_bar,eq:e}, as well as the relationship between \(\overline{\mathcal{E}}_{k}\) and \(\overline{\mathcal{E}}_{k, k + 1}\) defined in \Cref{eq:eqv_ebar}.

\begin{Lemma}
We have that
\begin{equation}
\label{eq:cond_ineq}
    \sup_{\overline{\mathbb{Q}} \in \overline{\mathcal{M}}} \mathbb{E}^{\overline{\mathbb{Q}}} [\Phi] \leq \overline{\mathcal{E}}^{0}(\Phi), \quad \Phi \in \overline{\Upsilon}.
\end{equation}
\label{lm:cond_ineq}
\end{Lemma}

\begin{proof}
We first rewrite \(\overline{\mathcal{E}}^{0}\) as \(\tilde{\mathcal{E}}^{0}\) below.
Let
\[
        \overline{\mathcal{G}}_k^- := \mathcal{G}_{k} \otimes \mathcal{B}(\mathcal{C}_{k - 1}) \subset \overline{\mathcal{G}}_k := \mathcal{G}_{k} \otimes \mathcal{B}(\mathcal{C}_k) \subset \overline{\mathcal{F}}_k,
    \]
\[
    \overline{\mathcal{M}}_k^-(\omega) := \left\{ \overline{\mathbb{Q}} \lll \overline{\mathcal{P}}: \overline{\mathbb{Q}} \left[ [\bar{\omega}]_{\overline{\mathcal{G}}_k^-}  \right] = 1 ~\text{and}~ \mathbb{E}^{\overline{\mathbb{Q}}} \left[ \Delta S_n \mid \overline{\mathcal{F}}_{n - 1} \right] = 0 ~\forall n \in \left\{ k + 1, \dots, N \right\} \right\},
    \]
which is essentially the same as \(\overline{\mathcal{M}}_{k}(\bar{\omega})\) defined in \Cref{eq:mk_bar}, with \([\bar{\omega}]_{\overline{\mathcal{G}}_{k}^-}\) defined as in \Cref{eq:om_atom}.
However, for the consistency with \(\overline{\mathcal{G}}_{k}^{-}\), we use \(\overline{\mathcal{M}}_{k}^{-}\) in the rest of this proof.

Note that \(\overline{\mathcal{G}}_{k}^{-}\) is countably generated because \(\mathcal{C}_{k - 1} \subseteq \mathcal{A}^{k - 1}\) is second countable and hence its Borel \(\sigma\)-algebra is countably generated.
For \(\Phi \in \overline{\Upsilon}\), define the operators
\begin{align*}
    \tilde{\mathcal{E}}_{k}(\varphi)(\bar{\omega}) &:= \sup_{\overline{\mathbb{Q}} \in \overline{\mathcal{M}}_{k}^{-}(\bar{\omega})} \mathbb{E}^{\overline{\mathbb{Q}}} \left[\varphi \right], \quad k \leq N - 1, \\[6pt]
    \tilde{\mathcal{E}}_{N}(\varphi)(\bar{\omega}) &:= \sup_{c' \in \mathcal{C}_{1}}\varphi(\bar{\omega}, c').
\end{align*}
Denote \(\overline{\mathcal{E}}^{k}(\cdot) := \overline{\mathcal{E}}_{k} \circ \dots \circ \overline{\mathcal{E}}_{N}(\cdot)\) and 
\(\tilde{\mathcal{E}}^{k}(\cdot) := \tilde{\mathcal{E}}_{k} \circ \dots \circ \tilde{\mathcal{E}}_{N}(\cdot)\). 
We claim that 
\begin{equation}
\label{eq:ebt_equiv}
    \overline{\mathcal{E}}^{k}(\Phi)(\bar{\omega}) = \tilde{\mathcal{E}}^{k}(\Phi)(\bar{\omega}), \quad 0 \leq k \leq N, \quad \Phi \in \overline{\Upsilon}.
\end{equation}

The fact that \(\overline{\mathcal{G}}_{k}^{-}\) is countably generated guarantees the existence of the regular conditional probabilities of any \(\mathbb{Q} \in \overline{\mathcal{M}}\) w.r.t. \(\overline{\mathcal{G}}_{k}^{-}\), denoted as \(\mathbb{Q}_{\bar{\omega}}\), 
which satisfies \linebreak \(\mathbb{Q} \left(\left\{ \bar{\omega}: \mathbb{Q}_{\bar{\omega}} \in \overline{\mathcal{M}}_{k}^{-}(\bar{\omega}) \right\}\right) = 1\). 
Then we have \(\mathbb{E}^{\mathbb{Q}} \left[\Phi \mid \overline{\mathcal{F}}_{k}^{-} \right] = \mathbb{E}^{\mathbb{Q}} \left[\Phi \mid \overline{\mathcal{G}}_{k}^{-} \right] \leq \tilde{\mathcal{E}}_{k}(\Phi)\), \(\mathbb{Q}\)-a.s.,
which implies \Cref{eq:cond_ineq} by the tower property of conditional expectations and the definition of \(\tilde{\mathcal{E}}^{0}\).

Hence, it suffices to prove \Cref{eq:ebt_equiv}. 
For \(\bar{\omega} = (\omega, c) \in \overline{\Omega}_{k}\), take any \(\mathbb{Q} \in \mathcal{M}_{k}(\omega)\), then for \(c' \in \mathcal{C}_{N - k + 1}\), 
we have that \(c'' = (c, c')\) and \(\mathbb{Q} \otimes \delta_{c''} \in \overline{\mathcal{M}}_{k}^{-}(\bar{\omega})\) and \(\mathbb{Q} \otimes \delta_{c''} \left(\Omega \times \{ c'' \}\right) = 1\). 
It then follows that \(\overline{\mathcal{E}}_{k}(f)(\bar{\omega}) \leq \tilde{\mathcal{E}}_{k}(f)(\bar{\omega})\), for \(f \in \overline{\Upsilon}\).

For the other direction, 
for \(\bar{\omega} = (\omega, c) \in \overline{\Omega}_{k}\), take any \(\overline{\mathbb{Q}} \in \overline{\mathcal{M}}_{k}^{-}(\bar{\omega})\)
and consider its regular conditional probabilities w.r.t. \(\overline{\mathcal{G}}_{k - 1}\), denoted by \(\overline{\mathbb{Q}}_{c''}\), \(c'' = (c, c')\) and \(c' \in \mathcal{C}_{N - k + 1}\). 
Then \(\overline{\mathbb{Q}}_{c'' \mid \Omega} \in \mathcal{M}_{k}(\omega)\) and \(\overline{\mathbb{Q}}_{c''} \left( \{ \omega \} \times \{c'' \} \right) = 1\), for all \(c \in \mathcal{C}_{N - k + 1}\). 

Therefore, \(\overline{\mathcal{E}}_{k}(f)(\bar{\omega}) \geq \tilde{\mathcal{E}}_{k}(f)(\bar{\omega})\), for \(f \in \overline{\Upsilon}\).
The result then follows by observing that \(\tilde{\mathcal{E}}_{N}(\varphi)(\bar{\omega}) = \overline{\mathcal{E}}_{N}(\varphi)(\omega, c)\).
\end{proof}

\begin{proof}[\bf Proof of \Cref{prop:dpp_bar}]
By \Cref{lm:ebar_eps}, for any \(\epsilon > 0\)
\begin{align*}
    \overline{\mathcal{E}}^{0}(\Phi) &= \overline{\mathcal{E}}_{0} \circ \dots \circ \overline{\mathcal{E}}_{N}(\Phi) \leq \mathcal{E}^{0}(\Phi_{\chi^{*}}) + \varepsilon = \sup_{\mathbb{Q} \in \mathcal{M}} \mathbb{E}^{\mathbb{Q}} \left[\Phi_{\chi^{*}} \right] + \varepsilon \\[6pt]
    & \leq \sup_{\mathbb{Q} \in \mathcal{M}} \sup_{\chi \in \mathcal{D}} \mathbb{E}^{\mathbb{Q}} \left[\Phi_{\chi} \right] + \varepsilon \leq \sup_{\overline{\mathbb{Q}} \in \overline{\mathcal{M}}} \mathbb{E}^{\overline{\mathbb{Q}}} \left[\Phi \right] + \varepsilon.
\end{align*}

The last inequality is because the set \(\overline{\mathcal{M}}\) is larger than the set of 
all push-forward measures induced by \(\omega \mapsto (\omega, \chi(\omega, \cdot))\) 
and \(\mathbb{Q} \in \mathcal{M}\) for all \(\chi \in \mathcal{D}\).
In addition to all feasible action sequences in \(\mathcal{D}\), it also takes into consideration all possible plans in \(\mathcal{C}\).
Since \(\varepsilon\) is arbitrary, we obtain the inequality
\[
        \overline{\mathcal{E}}^0(\Phi) \le \sup_{\mathbb{Q} \in \mathcal{M}} \sup_{\chi \in \mathcal{D}} \mathbb{E}^{\mathbb{Q}} \left[\Phi_{\chi} \right] \le \sup_{\overline{\mathbb{Q}} \in \overline{\mathcal{M}}} \mathbb{E}^{\overline{\mathbb{Q}}} \left[\Phi \right].
    \]
Combined with the inequality in \Cref{eq:cond_ineq} as proved in \Cref{lm:cond_ineq}, we have that
\[
        \overline{\mathcal{E}}^0(\Phi) = \sup_{\overline{\mathbb{Q}} \in \overline{\mathcal{M}}} \mathbb{E}^{\overline{\mathbb{Q}}} \left[\Phi \right] = \sup_{\mathbb{Q} \in \mathcal{M}} \sup_{\chi \in \mathcal{D}} \mathbb{E}^{\mathbb{Q}} \left[\Phi_{\chi} \right].
    \]
\end{proof}

The local no-arbitrage condition \(NA\left(\mathcal{P}_{k, k + 1}(\omega)\right)\) introduced in \cite{bouchard_2015_arbit_nondo} states that, 
given a fixed \(\omega \in \Omega_{k}\), \(\Delta S_{k + 1}(\omega, \cdot) := S_{k + 1}(\omega, \cdot) - S_{k}(\omega)\) can be considered as a random variable on \(\Omega_{1}\),
which represents a one-period market on \(\left(\Omega_{1}, \mathcal{B}(\Omega_{1})\right)\) endowed with a collection \(\mathcal{P}_{k, k + 1}(\omega)\) of probability measures.
Then \(NA\left(\mathcal{P}_{k, k + 1}(\omega)\right)\) denotes the no-arbitrage condition in such a one-period market; that is,
the condition \(NA(\mathcal{P}_{k, k + 1}(\omega))\) holds if for all \(H \in \mathbb{R}^{d}\),
\[
    H\Delta S_{k + 1} (\omega, \cdot) \ge 0 \quad  \mathcal{P}_{k, k + 1} (\omega)\text{-q.s.} \implies H\Delta S_{k + 1} (\omega, \cdot) = 0 \quad \mathcal{P}_{k, k + 1} (\omega)\text{-q.s.}
\]

Recall the definition of \(\overline{\Omega}_{k}^{+} := \Omega_{k} \times \mathcal{C}_{k}\), then by \Cref{lm:mrb_chi} and by Lemma 4.10 of \cite{bouchard_2015_arbit_nondo}, we obtain the following result.
\begin{Lemma}
Let \(f: \overline{\Omega}_{k + 1} \to \overline{\mathbb{R}}\) be upper semianalytic. Then \(\overline{\mathcal{E}}_{k}(f): \overline{\Omega}_{k} \to \overline{\mathbb{R}}\) is still upper semianalytic.
Moreover, there exists a universally measurable function \(y: \overline{\Omega}_{k}^{+} \to \mathbb{R}^{d}\) such that
\[
        \overline{\mathcal{E}}(f)(\bar{\omega}) + y(\bar{\omega}^+)\Delta S_{k + 1} (\omega, \cdot) \ge f(\omega, \cdot, c), \quad  \mathcal{P}_{k, k + 1} (\omega)\text{-q.s.}
    \]
for all \(\bar{\omega}^+ = (\omega, c) = (\bar{\omega}, c_k) \in \overline{\Omega}_k^+\) such that \(NA\left(\mathcal{P}_{k, k + 1}(\omega)\right)\) holds and \(f(\omega, \cdot, c) > -\infty\), \(\mathcal{P}_{k, k + 1}(\omega)\)-q.s.
\label{lm:slr_exist}
\end{Lemma}

We are ready for the proof of \Cref{thm:dual_bar}.
\begin{proof}[\bf Proof of \Cref{thm:dual_bar}]
The weak duality as in \Cref{eq:weak_dual} is given by
\[
        \sup_{\overline{\mathbb{Q}} \in \overline{\mathcal{M}}} \mathbb{E}^{\overline{\mathbb{Q}}} [\Phi] \le \overline{\pi}^E (\Phi).
    \]
Without loss of generality, assume that \(\Phi\) is bounded from above, as by Lemma 6.4 of \cite{aksamit_2019_robus_prici} we have that
\(\lim_{n \to \infty} \overline{\pi}^{E}(\Phi \wedge n) = \overline{\pi}^{E}(\Phi)\) (see also the proof of Theorem 3.4 of \cite{bouchard_2015_arbit_nondo}) and that
\(\sup_{\overline{\mathbb{Q}} \in \overline{\mathcal{M}}} \mathbb{E}^{\overline{\mathbb{Q}}} \left[\Phi \right] = \lim_{n \to \infty} \sup_{\overline{\mathbb{Q}} \in \overline{\mathcal{M}}} \mathbb{E}^{\overline{\mathbb{Q}}} \left[\Phi \wedge n \right]\), by the monotone convergence theorem.

When \(\Phi\) is bounded from above, by \Cref{prop:dpp_bar}, it is enough to prove that there exists some \(\overline{H} \in \overline{\mathcal{H}}\) such that
\[
        \overline{\mathcal{E}}^0(\Phi) + \left(\overline{H} \circ S \right)_N \ge \Phi \quad  \overline{\mathcal{P}}\text{-q.s.}
    \]

By \Cref{lm:mtg_intg}, we have that \(\overline{\mathcal{E}}^{k}(\Phi)(\bar{\omega}) > -\infty\) for all \(\bar{\omega} \in \overline{\Omega}_{k}\). 
Recall that the only difference between \(\overline{\Omega}_{k}\) and \(\overline{\Omega}_{k}^+\) is the one-step action at time \(k\).
Then by \Cref{lm:slr_exist}, there exist a universally measurable function \(y^{k}: \overline{\Omega}_{k}^{+} \to \mathbb{R}^{d}\) such that
\[
        y^k(\bar{\omega}^+)\Delta S_{k + 1} (\omega, \cdot) \ge \overline{\mathcal{E}}^{k + 1} (\Phi)(\omega, \cdot, c) - \overline{\mathcal{E}}^k(\bar{\omega}), \quad  c' \in \mathcal{C}_{N - k + 1}, \quad \mathcal{P}_{k, k + 1}(\omega)\text{-q.s.},
    \]
for all \(\bar{\omega} \in \overline{\Omega}_{k}^{+}\) such that \(NA(\mathcal{P}_{k, k + 1}(\omega))\) holds.

It follows that
\begin{align*}
    \sum _{k = 0} ^{N - 1} \overline{H}_{k + 1} \Delta S_{k + 1} &\ge \sum _{k = 0} ^{N - 1} \left( \overline{\mathcal{E}}^{k + 1} (\Phi) - \overline{\mathcal{E}}^k \right) \\[6pt]
    &= \overline{\mathcal{E}}_{N}(\Phi) - \overline{\mathcal{E}}^{0}(\Phi) \ge \Phi - \overline{\mathcal{E}}^{0}(\Phi), \quad  \mathcal{P}\text{-q.s.,}
\end{align*}
where \(\overline{H}_{k + 1}(\bar{\omega}) := y^{k}([\bar{\omega}]_{k}^{+})\).

Hence, \(\overline{H}\) is an optimal superhedging strategy for \(\Phi\) being bounded above. 
The existence of the optimal superhedging strategy for general \(\Phi\) is then a consequence of Lemma 6.4 in \cite{aksamit_2019_robus_prici}.
\end{proof}

\section{Duality with information delay}
\label{sec:info_delay}
\subsection{The delay setup}
\label{sec:org662cd64}
Let us consider the setup where the information on the action chosen by the buyer at time \(k\) is only revealed to the seller up to \(l \le N\) periods later.  At maturity \(N\), the seller would have known the true sequence of actions chosen by the buyer.

To this end, denote \(\widehat{\mathcal{C}}:= \widehat{\mathcal{A}}^{N + 1}\) to be the space of true action sequences, where \(\widehat{\mathcal{A}}\) is the space of true actions at each time. Let \(\mathcal{A} := \left( \widehat{\mathcal{A}} \cup \left\{ null \right\} \right)^{N + 1}\), and let \(\mathcal{C} := \mathcal{A}^{N + 1}\) be the space of observed actions at each time. We explicitly include a \(null\) element to take care of invalid/conflicting actions, and we will talk more about this later (for example, \Cref{exmp:bad}).

More specifically, for a sequence of true actions \(\hat{c} = \left( \hat{c}_{0}, \hat{c}_{1}, \dots, \hat{c}_{N} \right)\), the corresponding observed sequence of actions \(c = \left( c_{0}, c_{1}, \dots, c_{N} \right)\), with \(c_{t} = \left( c_{t, 0}, c_{t, 1}, \dots, c_{t, N} \right)\) the observation at time \(t\), is given by
\begin{equation}
\label{eq:obs_matrix}
\begin{cases}
  c_{t, k} = \hat{c}_{k} \in \widehat{\mathcal{A}}, & \hat{c}_{k} ~\text{is observed exactly at time}~ t, \\[6pt]
  c_{t, k} = null, &\text{otherwise},
\end{cases}
\end{equation}
for \(t = 0, \dots, T\). That is, \(c_{t, k}\) represents the information observed at time \(t\) for the true action \(\hat{c}_{k}\) occurred at time \(k\); if the information of \(\hat{c}_{k}\) is not available at time \(t\) or has been obtained before time \(t\), we set \(c_{t, k}\) to null.

\begin{Example}
\label{exmp:one}
Consider a \( 2\)-period case with true action space at each time \( \widehat{\mathcal{A}} = \left\{ L, R \right\}\). We have \( \mathcal{A} := \left\{ L, R, null \right\}^{3} \) and \( \mathcal{C} := \mathcal{A}^{3}\). Then the observed sequence
\[
c = \left( c_{0}, c_{1}, c_{2} \right) ~\text{with}~ c_{0} = \left( null, null, null \right), c_{1} = \left( R, null, null \right), c_{2} = \left( null, L, R \right)
\]
corresponds to the situation that \( R\) (resp. \( L, R\)) is the true action at time \( 0\) (resp. \( 1, 2\)) observed at time \( 1\) (resp. \( 2, 2\)); the observed sequence
\[
c' = \left( c'_{0}, c'_{1}, c'_{2} \right), ~\text{with}~ c'_{0} = \left( R, null, null \right), c'_{1} = \left( null, null, null \right), c'_{2} = \left( null, L, L \right)
\]
corresponds to the situation that \( R\) (resp. \( L, L\)) is the true action at time \( 0\) (resp. \( 1, 2\)) observed at time \( 0\) (resp., \( 2, 2\)).
\end{Example}

Fix \(l \le N\), and let \(\mathcal{C}^{v}\) be the space of observed action sequences in the variable delay case (i.e., the delay could take any value in \(\left\{ 0, 1, \dots, l \right\}\)). Let \(\mathcal{C}^{f}\) be the space of observed action sequences in the fixed-delay case where the delay is fixed to be \(l\). Obviously, \(\mathcal{C}^{f} \subseteq \mathcal{C}^{v} \subseteq \mathcal{C}\).

Moreover, for any \(c = \left( c_{0}, c_{1}, \dots, c_{N} \right) \in \mathcal{C}^{v}\) with \(c_{t} = \left( c_{t, 0}, c_{t, 1}, \dots, c_{t, N} \right)\), for any \(k\) there is exactly one \(t\) such that \(c_{t, k} \in \widehat{\mathcal{A}}\), meaning that the true action at time \(k\) is only observed once.

\begin{Example}
Consider the observed sequences \( c, c'\) from \Cref{exmp:one}. When \( l = 1\), we have \( c \in \mathcal{C}^{f} \subseteq \mathcal{C}^{v}\) and \( c' \in \mathcal{C}^{v} \setminus \mathcal{C}^{f}\).
\end{Example}

Define \(\xi^{v} : \mathcal{C}^{v} \rightarrow \widehat{\mathcal{A}}^{N + 1}\) (resp. \(\xi^{f} : \mathcal{C}^{f} \rightarrow \widehat{\mathcal{A}}^{N + 1}\)) to be the mapping from the observed action sequence space to the true action sequence space via \Cref{eq:obs_matrix} in an obvious way; that is,
\[
\xi^{v} = \left( \xi^{v}_{0}, \dots, \xi^{v}_{N} \right) ~\text{with}~ \xi^{v}_{k}(c) = \hat{c}_{k}, \quad  c \in \mathcal{C},
\]
where \(\hat{c}_{k} \in \widehat{\mathcal{A}}\) is the true action at time \(k\). It is easy to see that \(\xi^{v}\) (resp. \(\xi^{f}\)) is surjective (resp. bijective) on \(\mathcal{C}^{v}\) (resp. \(\mathcal{C}^{f}\)).

We further extend the domain of \(\xi^{v}(\cdot)\) (resp. \(\xi^{f}(\cdot)\)) to \(\mathcal{C}\) and set \(\xi^{v}(c)\) (resp. \(\xi^{f}(c)\)) to a null sequence if \(c \not\in \mathcal{C}^{v}\) (resp. \(c \not\in \mathcal{C}^{f}\)), which will trigger the payoff to be set to \(-\infty\) as we will see later.

\begin{Example}
\label{exmp:bad}
Consider again the setup in \Cref{exmp:one} and let \( l = 1\). We have
\[
\xi^{v}(c) = \xi^{f}(c) = \left( R, L, R \right), \quad \xi^{v}(c') = \left( R, L, L \right) \quad \text{and}~~ \xi^{f}(c) = ~\text{null sequence}. 
\]
For the observed sequence
\[
c'' = \left( c''_{0}, c''_{1}, c''_{2}\right) ~\text{with}~ c''_{0} = \left( null, null, null \right), c''_{1} = \left( R, R, null \right), c''_{2} \left( null, L, L \right)
\]
since \( c'' \not\in \mathcal{C}^{v}\) (the true action at time \( 2\) cannot be observed more than once and it is also conflicting), we have
\[
\xi^{v}(c'') = \xi^{f}(c'') = ~\text{null sequence}.
\]

\end{Example}

Let \(\Phi: \Omega \times \widehat{\mathcal{A}}^{N + 1} \to \overline{\mathbb{R}}\) be upper semianalytic, representing the payoff function of an option to be superhedged. Note that \(\Phi\) really is a function of the true action sequence instead of the observed one. We further set
\begin{equation}
\label{eq:pay_null}
\Phi(\cdot, \text{null sequence}) := -\infty
\end{equation}
to count for invalid observed sequences. Recall the set of dynamic trading strategies
defined in \Cref{eq:strat}. We define the superhedging price for the variable-delay case:
\begin{equation}
\label{eq:gen_price}
\pi^{v}(\Phi) := \inf \left\{x: \exists H \in \mathcal{H} ~\text{s.t.}~ x + \left( H(\cdot, c) \circ S \right)_N \ge \Phi(\cdot, \xi^{v}(c)) ~ \mathcal{P}\text{-q.s.}, \forall c \in \mathcal{C} \right\},
\end{equation}
as well as for fixed-delay case:
\begin{equation}
\label{eq:fixed_price}
\pi^{f}(\Phi) := \inf \left\{x: \exists H \in \mathcal{H} ~\text{s.t.}~ x + \left( H(\cdot, c) \circ S \right)_N \ge \Phi(\cdot, \xi^{f}(c)) ~ \mathcal{P}\text{-q.s.}, \forall c \in \mathcal{C} \right\}.
\end{equation}
Recall the space of admissible sequences of actions defined in \Cref{eq:def_d}.
With a bit abuse of notation, we write \(\Phi_{\xi^{v}(\chi)} = \Phi \left( \cdot, \xi^{v}(\chi(\cdot, \cdot)) \right)\), and present our result on duality with information delay as the following theorem.

\subsection{Main results}
\label{sec:orgd22e649}
\begin{Theorem}
We have the following dualities
\begin{equation}
\label{eq:pred_dual}
    \pi^{v}(\Phi) = \sup_{\mathbb{Q} \in \mathcal{M}} \sup_{\chi \in \mathcal{D}} \mathbb{E}^{\mathbb{Q}} \left[\Phi_{\xi^{v}( \chi)} \right] = \pi^{f}(\Phi) = \sup_{\mathbb{Q} \in \mathcal{M}}  \sup_{\chi \in \mathcal{D}} \mathbb{E}^{\mathbb{Q}} \left[\Phi_{\xi^{f} (\chi)} \right] = \sup_{\mathbb{Q} \in \mathcal{M}} \sup_{\chi^a \in \mathcal{D}^a} \mathbb{E}^{\mathbb{Q}} \left[\Phi_{\chi^a} \right].
\end{equation}
where \(\Phi_{\chi^a} = \Phi\left(\cdot, \chi^a(\cdot, \cdot) \right)\) and
\[
\mathcal{D}^a := \left\{ \chi^a : \Omega \times \mathbb{T} \to \widehat{\mathcal{A}} \mid \chi^a(\cdot, k) ~\text{is \( \mathcal{F}_{(k + l) \wedge N} \)-measurable}, ~k = 0, ..., N \right\}.
\]
\label{thm:info_dual}
\end{Theorem}

\begin{Remark}
\Cref{thm:info_dual} says that the superhedging price in the variable-delay situation is the same as that for the fixed-delay,
which should not come as a surprise since we are in the robust framework and it makes sense to hedge against the worst case scenario.
The more interesting part is when there is an information delay, the superhedging price increases, hence breaking the duality.
But then, the duality can be restored, by fictitiously allowing the buyer the ability to look into the future as far as the maximum delay, and thus increasing the worst case model price.
\label{rm:info_dual}
\end{Remark}

\begin{Remark}
In \cite[][Section 3.1]{bayraktar_2017_hedgi_ameri}, it is shown that if the seller of the hedged American option does not adjust her trading strategy according to the realized exercise time (which is equivalent to maximum delay periods \(l = N\)),
then the superhedging price would be the same as the supremum of the expectation of the pathwise maximum of the Americaion option over martingale measures (which is aligned with the case that the option holder can look into the future for \(l = N\) periods).
Therefore, the result in \cite[][Section 3.1]{bayraktar_2017_hedgi_ameri} can be thought of as a special case of Theorem 4.3. 
\end{Remark}

\begin{Remark}
As pointed out in \Cref{rm:dym}, when statically traded European options \(\textsl{g}\) are included, we may establish the duality as follows by a dynamic extension such that the European options can also be dynamically traded,
\begin{align*}
\label{eq:enlarge}
\pi^{v}_{\textsl{g}} (\Phi) = \widehat{\pi}^{v} (\Phi) = \sup_{\widehat{\mathbb{Q}} \in \widehat{\mathcal{M}}} \sup_{\widehat{\chi} \in \widehat{\mathcal{D}}} \mathbb{E}^{\widehat{\mathbb{Q}}} \left[\Phi_{\xi^{v} \left( \widehat{\chi} \right)} \right] &= \pi^{f}_{\textsl{g}} (\Phi) = \widehat{\pi}^{f} (\Phi) \\[6pt]
&= \sup_{\widehat{\mathbb{Q}} \in \widehat{\mathcal{M}}} \sup_{\widehat{\chi} \in \widehat{\mathcal{D}}} \mathbb{E}^{\widehat{\mathbb{Q}}} \left[\Phi_{\xi^{f} \left( \widehat{\chi} \right)} \right] = \sup_{\widehat{\mathbb{Q}} \in \widehat{\mathcal{M}}} \sup_{\widehat{\chi}^a \in \widehat{\mathcal{D}}^a} \mathbb{E}^{\widehat{\mathbb{Q}}} \left[\Phi_{\widehat{\chi}^a} \right].
\end{align*}
\end{Remark}

\begin{proof}[\bf Proof of \Cref{thm:info_dual}]
Since \(\Phi(\cdot, \xi^{v}(c)) \ge  \Phi(\cdot, \xi^{f} (c))\) for any \(c \in \mathcal{C}\), we must have \(\pi^{f}(\Phi) \le \pi^{v}(\Phi)\).
On the other hand, let \(x \in \mathbb{R}\) and \(H \in \mathcal{H}\) be such that
\[
x + \left( H(\cdot, c) \circ S \right)_{N} \ge \Phi(\cdot, \xi^{f}(c)), ~\mathcal{P}\text{-q.s.,} ~\forall c \in \mathcal{C}^{f} \quad \text{(and thus}~ \forall c \in \mathcal{C} ~\text{due to \eqref{eq:pay_null}}).
\]
Let \(\left( \xi^{f} \right)^{-1} : \widehat{\mathcal{C}} \rightarrow \mathcal{C}^{f}\) be the inverse function of \(\xi^{f}\). Define \(H' \in \mathcal{H}\) by
\begin{align*}
H_t'(\omega, c)=
\begin{cases}
H_t(\omega, \left( \xi^{f} \right)^{-1} \left( \xi^{v}(c) \right)),& {\exists c'\in \mathcal{C}^v} ~\text{s.t.}~ c_k'=c_k, k=0, \dots, t,\\
0,&\text{otherwise.}
\end{cases}
\end{align*}

The idea is that, in a variable-delay situation, the hedger could make it through by always mimicking the strategies of a hedger in a fixed-delay situation who is exposed to less information. Then we have that
\[
x + \left( H'(\cdot, c) \circ S \right)_{N} \ge \Phi(\cdot, \xi^{v}(c)), ~\mathcal{P}\text{-q.s.,} ~\forall c \in \mathcal{C}^{v} \quad \text{(and thus}~ \forall c \in \mathcal{C} ~\text{due to \eqref{eq:pay_null}}~).
\]
This implies that \(\pi^{f}(\Phi) \ge \pi^{v}(\Phi)\) and thus \(\pi^{f}(\Phi) = \pi^{v}(\Phi)\). By \Cref{thm:dual_bar}, we have
\begin{equation}
\label{eq:info_dual}
\pi^{v} (\Phi) = \sup_{\mathbb{Q} \in \mathcal{M}} \sup_{\chi \in \mathcal{D}}  \mathbb{E}^{\mathbb{Q}} \left[ \Phi_{\xi^{v}(\chi)} \right] = \pi^{f} (\Phi) = \sup_{\mathbb{Q} \in \mathcal{M}} \sup_{\chi \in \mathcal{D}} \mathbb{E}^{\mathbb{Q}} \left[ \Phi_{\xi^{f}(\chi)} \right].
\end{equation}

Alternatively, by \Cref{thm:dual_bar} the first and third equality hold in \Cref{eq:info_dual}. Since \(\Phi(\cdot, \xi^{v}(c)) \ge \Phi(\cdot, \xi^{f}(c))\) for any \(c \in \mathcal{C}\), we have
\[
\sup_{\mathbb{Q} \in \mathcal{M}} \sup_{\chi \in \mathcal{D}} \mathbb{E}^{\mathbb{Q}} \left[ \Phi_{\xi^{v}(\chi)} \right] \ge \sup_{\mathbb{Q} \in \mathcal{M}} \sup_{\chi \in \mathcal{D}} \mathbb{E}^{\mathbb{Q}} \left[ \Phi_{\xi^{f}(\chi)} \right].
\]
On the other hand, for any \(\chi \in \mathcal{D}\) with \(\xi^{v} \left( \chi(\cdot) \right) \in \mathcal{C}^{v} ~\mathcal{P}\)-q.s., since \(\xi^{f}\) is bijective on \(\mathcal{C}^{f}\), there exists a \(\tilde{\chi} \in \mathcal{D}\) such that
\[
\xi^{v} \left( \chi(\omega) \right) = \xi^{f} \left( \tilde{\chi} (\omega) \right), \quad \mathcal{P}\text{-q.s.}
\]
This implies
\[
\sup_{\mathbb{Q} \in \mathcal{M}} \sup_{\chi \in \mathcal{D}} \mathbb{E}^{\mathbb{Q}} \left[ \Phi_{\xi^{v} (\chi)} \right] \le \sup_{\mathbb{Q} \in \mathcal{M}} \sup_{\chi \in \mathcal{D}} \mathbb{E}^{\mathbb{Q}} \left[ \Phi_{\xi^{f} (\chi)} \right]
\]
and thus \Cref{eq:info_dual}.

It remains to show the last equality in \Cref{eq:pred_dual}. Take \(\chi \in \mathcal{D}\), and without loss of generality we assume \(\chi(\cdot) \in \mathcal{C}^{f} ~\mathcal{P}\)-q.s., for otherwise \(\mathbb{E}^{\mathbb{Q}} \left[ \Phi_{\xi^{f}(\chi)} \right] = -\infty\) for \(\mathbb{Q} \in \mathcal{M}\). Then it is easy to see that \(\xi^{v}_{k} \left( \chi(\cdot) \right)\) is \(\mathcal{F}_{\left( k + l \right) \wedge N}\)-measurable (up to a \(\mathcal{P}\)-null set) for \(k = 0, \dots, N\).
Therefore,
\[
\sup_{\mathbb{Q} \in \mathcal{M}} \sup_{\chi \in \mathcal{D}} \mathbb{E}^{\mathbb{Q}} \left[ \Phi_{\xi^{f}(\chi)} \right] \le \sup_{\mathbb{Q} \in \mathcal{M}} \sup_{\chi^{a} \in \mathcal{D}^{a}} \mathbb{E}^{\mathbb{Q}} \left[ \Phi_{\chi^{a}} \right].
\]

Conversely, let \(\chi^{a} \in \mathcal{D}^{a}\). As \(\xi^{f}\) is bijective on \(\mathcal{C}^{f}\), we may take \(\chi \in \mathcal{D}\) such that \(\xi^f (\chi) = \chi^{a}\), i.e., \(\chi_{k}(\cdot) = \left( \chi_{k, 0}(\cdot), \dots, \chi_{k, N}(\cdot) \right) = \left( (\xi^{f})^{-1} (\chi^{a}) \right)_{k}\) with
\begin{align*}
\chi_{k, t}(\cdot)=
\begin{cases}
\chi^a_t(\cdot), &k = (t + l)\wedge N,\\
null,&\text{otherwise},
\end{cases}
\end{align*}

for \(k, t = 0, \dots, N\). Note that since \(\chi^{a}_{t}\) is \(\mathcal{F}_{\left( t + l \right) \wedge N}\)-measurable for \(k \in \mathbb{T}\), \(\chi_{k}\) is indeed \(\mathcal{F}_{k}\)-measurable for \(k \in \mathbb{T}\) and thus we indeed have \(\chi \in \mathcal{D}\). Consequently,
\[
\sup_{\mathbb{Q} \in \mathcal{M}} \sup_{\chi \in \mathcal{D}} \mathbb{E}^{\mathbb{Q}} \left[ \Phi_{\xi^{f}(\chi)} \right] \ge \sup_{\mathbb{Q} \in \mathcal{M}} \sup_{\chi^{a} \in \mathcal{D}^{a}} \mathbb{E}^{\mathbb{Q}} \left[ \Phi_{\chi^{a}} \right],
\]
and the proof is complete.
\end{proof}

\begin{Remark}
The proof of \Cref{thm:info_dual} relies on the measurability of  \(\xi^{f} : \mathcal{C} \to \widehat{\mathcal{C}}\), and here is a quick note on why it would not be of concern. Rewrite \(\xi^{f} := \left( \xi^{f}_0, \dots, \xi^{f}_N \right)\), where \(\xi^{f}_k : \mathcal{C} \to \widehat{\mathcal{A}}\). It is easy to see that for any \(A \in \mathcal{F}^{\widehat{\mathcal{A}}}\), we have that  \(\left( \xi^{f} \right)^{-1}_k \left( A \right) = \textit{null action} \times A \times \dots \times \textit{null action}\) which belongs to \(\mathcal{F}^{\mathcal{C}}\). Therefore, \(\xi^{f}_k\) is measurable for all \(k \in \left\{ 0, 1, \dots, N \right\}\), and so is \(\xi^{f}\).
For the variable-delay, we can write \(\left( \xi^{v}_{k}\right)^{-1} (A) = \bigcup_{j = 0}^l \left( \xi^{f, j}_{k} \right)^{-1} (A)\) where \(j \in \left\{ 0, \dots, l \right\}\) is the length of the variable-delay period, which is also in \(\mathcal{F}^{\mathcal{C}}\).
\end{Remark}

\begin{Remark}
It is worth pointing out that (in the variable-delay situation) the delay could be a choice of the buyer, or it could be due to some randomness from the external world, for instance, a communication system breakdown, etc.
\end{Remark}

\subsection{The delayed Snell envelope}
\label{sec:orgad27f4a}
The Snell envelope, first introduced in \cite{snell_1952_applic_martin}, is an important tool used in the valuation of American options.
In this section, we develop a similar notion in the context of an information delay on the buyer's decision to exercise an American option, in a nonrobust complete market setup.

Consider the following motivating example to see how the delay would cause the American option price to go up.
\begin{Example}
\label{exmp_snell}
Consider the case with one-American option. Let \( N\) be the maturity date and \( l < N\) be the number of maximum delay periods. The true action space is given by \( \widehat{\mathcal{A}} := \left\{ stop, continue \right\}\). The collection of all possible action plans would be \( \widehat{\mathcal{C}} := \widehat{\mathcal{A}}^{N + 1} \), while the observed action space at each time would be of the form \( \mathcal{A} := \left(\widehat{\mathcal{A}} \cup \left\{ null \right\} \right)^{N + 1}\).

For instance, let \( N =  14\) and \( l = 3\). If the buyer chooses to exercise at time \( 2\) with a delay period of \( 2\), then only when \( 2\) periods later at time \( 4\) would the option seller be able to observe that the option has been exercised already at time \( 2\).

To see how the information delay could cause the model price to go up, consider the following non-robust \( 2\)-period American option example:
\begin{figure}[H]
\label{fig:eg_am}
\centering
      \begin{tikzpicture}
          \draw[] (0, 0) -- (2, 2) node [midway, above, sloped] (TextNode) {\color{green}{\( 1/2\)}};
          \draw[] (0, 0) -- (2, -2) node [midway, below, sloped] (TextNode) {\color{green}{\( 1/2\)}};
          \draw[] (2, 2) -- (4, 3) node [midway, above, sloped] (TextNode) {\color{green}{\( 1/2\)}};
          \draw[] (2, 2) -- (4, 1) node [midway, below, sloped] (TextNode) {\color{green}{\( 1/2\)}};
          \draw[] (2, -2) -- (4, -1) node [midway, above, sloped] (TextNode) {\color{green}{\( 1/2\)}};
          \draw[] (2, -2) -- (4, -3) node [midway, below, sloped] (TextNode) {\color{green}{\( 1/2\)}};

          \draw [fill] (2, 2) circle [radius = .1] node[left]{(4; {\color{blue} 1})};
          \draw [fill] (2, -2) circle [radius = .1] node[left]{(2; {\color{blue} 2})};
          \draw [fill] (4, 3) circle [radius = .1] node[right]{(6; {\color{blue} 4})};
          \draw [fill] (4, 1) circle [radius = .1] node[right]{(2; {\color{blue} 2})};
          \draw [fill] (4, -1) circle [radius = .1] node[right]{(4; {\color{blue} 1})};
          \draw [fill] (4, -3) circle [radius = .1] node[right]{(0; {\color{blue} 6})};

          \draw [fill] (0, 0) circle [radius = .1] node [left] {(3; {\color{blue} 0})};
      \end{tikzpicture}    
      \caption{{\it \( 2\)-period binomial tree for an American option.} The labels at each node is a tuple of the stock price and {\color{blue} the payoff of the American option},  respectively.}
\end{figure}
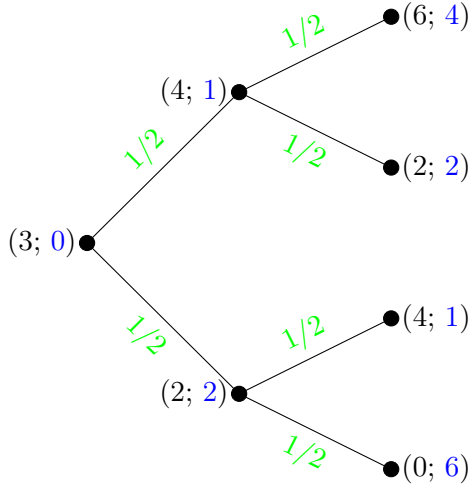
In the no-delay situation, the model price is given by
\[
\sup_{\tau} \mathbb{E}^{\mathbb{Q}} [\Phi_{\tau}] =  \frac{1}{2} \left( 2 \vee 1 + 3 \vee 2 \right) \vee 0 = \frac{5}{2},
\]
while in the \( 1\)-period fixed delay setup, by considering at time \( 2\) the immediate payoff versus the payoff that one should have got if exercising the option \( 1\) period earlier, we find that
\[
\sup_{\tau^a} \mathbb{E}^{\mathbb{Q}} [\Phi_{\tau^a}] =  \frac{1}{2} \left( \frac{1}{2} \left( 4 \vee 1 + 2 \vee 1 \right) \vee 1 + \frac{1}{2} \left( 1 \vee 2 + 6 \vee 2 \right) \vee 2 \right) = \frac{12}{4}.
\]
\end{Example}

Recall that \(l \in \mathbb{T}\) is the maximum delay, a fixed number, and we introduce the following set of random times
\[
\mathcal{T}^{-l}_{s} \coloneqq \left\{ \tau: \tau \in [(s - l) \vee 0,  N] ~\text{and}~ \left\{ \tau \le (s - l) \vee 0 \right\} ~\text{is}~ \mathcal{F}_{s}\text{-measurable} \right\},  \quad ~\text{for any}~ s \in \mathbb{T}.
\]
Then the Snell envelop from the seller's point of view is given by
\[
J_{k}^{-l} = \sup_{\tau \in \mathcal{T}_{k}^{-l}} \mathbb{E}^{\mathbb{Q}} \left[ \Phi_{\tau} \mid \mathcal{F}_{k} \right],
\]
the intuition being that given \(\mathcal{F}_{k}\) at time \(k\), the seller, dreading the buyer's ability to revise her decisions in the past,
assumes the worst-case scenario where the option has not been exercised up to \(l\) units of time ago.

On the other hand, let \(\mathcal{K}_{s} \coloneqq \mathcal{F}_{(s + l) \wedge N}\) and \(\mathbb{K} \coloneqq (\mathcal{K}_{s})_{s = 0, 1, \dots, N}\) and we have
\[
\mathcal{T}_{s}^{+ l} \coloneqq \left\{ \tau: \tau \in [s, N] ~\text{and}~ \tau ~\text{is a}~ \mathbb{K}\text{-stopping time} \right\}, \quad ~\text{for any}~ s \in \mathbb{T}.
\]
From the buyer's perspective, the Snell envelope is of the following form
\[
J_{k}^{+ l} \coloneqq J_{k + l}^{- l} = \sup_{\tau \in \mathcal{T}_{k}^{+ l}} \mathbb{E}^{\mathbb{Q}} \left[ \Phi_{\tau} \mid \mathcal{K}_{k} \right],
\]
which takes into consideration the fictitious power that she has to look into the future for up to \(l\) periods.

It is easy to see that \(J_{k}^{- l}\) coincides with \(J_{k}^{+ l}\) by a change of variable.
Proof of \(J_{k}^{+ l}\) being a \((\mathbb{Q}, \mathbb{K})\)-supermartingale and the smallest supermartingale dominating \(\Phi\) can be found in, for example, \cite[][Appendix I]{della_2011_proba_poten}, \cite[][Proposition 3.2]{guo_2014_arbit_pricin}

Denote \(\mathcal{T}_{0}^{+ l}\) by \(\mathcal{T}^{a}\) as introduced in the previous sections, we have
\[
J_{0}^{+ l} = \sup_{\tau \in \mathcal{T}^{a}} \mathbb{E}^{\mathbb{Q}} \left[ \Phi_{\tau} \right] = \bar{\pi}(\Phi),
\]
by \Cref{thm:info_dual}, and by noticing that the nonrobust complete market setup reduces both \(\mathcal{P}\) and \(\mathcal{M}\) to singletons.

Then the Doob-Meyer decomposition theorem can be applied, yielding
\[
J_{k}^{+ l} = M_{k} - C_{k},
\]
where \(M\) is a \((\mathbb{Q}, \mathbb{K})\)-martingale and \(C\) a nondecreasing, integrable and \(\mathbb{K}\)-predictable process with \(C_{0} = 0\).

Consider \(N_{k} = M_{k - l}\) and we see that \(N\) is a \((\mathbb{Q}, \mathbb{F})\)-martingale, which guarantees the existence of some hedging strategy, say \(\tilde{H}\).
However, note that the first piece of information regarding the buyer's decision only comes in at time \(l\),
hence it is reasonable to let \(\tilde{H}_{0}, \dots, \tilde{H}_{l - 1}\) follow the hedging strategy of the European option \(\Phi\) terminating at time \(l - 1\).

\begin{Remark}
Note that \(\mathcal{T}^{+ l}_{s}\) can be obtained via a change variable on \(\mathcal{T}^{- l}_{s}\).
Intuitively, they have described the option buyer's fictitious super power in the information delay setup from two different perspectives;
that is, the power to rectify past decisions and the power to make decisions based on future information. 
\end{Remark}

\section{Examples}
\label{sec:exmp}
Apart from the American option in an information delay setup, as given in \Cref{exmp_snell}, we also present the following few examples where our model is applicable.

\subsection{Multiple American options with constraints}
\label{sec:exmp_multi}
Say we want to sell \(m \in \mathbb{N}\) American options,
how should we price it and do we still have a nice duality result like we used to?

Without loss of generality, we exclude exercise at time \(0\). Let \(\mathbb{T} := \left\{ 1, \dots, N \right\}\), which also serves as our action space; that is, we have that \(\mathcal{A} := \mathbb{T}\).
Hence, the collection of all possible stopping plans is denoted as \(\mathcal{C} := \mathbb{T}^{m}\), and the enlarged space in this case can be defined as \(\overline{\Omega} := \Omega \times \mathcal{C}\), with \(T: \overline{\Omega} \to \mathbb{T}^{m}\) given by \(T(\bar{\omega}) = \left(\theta_{1}, \dots, \theta_{m} \right)\).
An element of \(\overline{\Omega}\) looks like \(\bar{\omega} = \left(\omega, \theta_{1}, \dots, \theta_{m} \right)\). The filtration is given by \(\overline{\mathbb{F}} := \left(\overline{\mathcal{F}}_{k} \right)_{k = 0, \dots, N}\) with \(\overline{\mathcal{F}}_{k} = \mathcal{F}_{k} \otimes \vartheta_{k}^{m}\) and \(\vartheta_{k} = \sigma \left(T \wedge (k + 1) \right)\) to ensure non-anticipativity.

An easy adaptation of our previous arguments, coupled with a random set at each time step to keep track of the American options that are still active, gives us the desired duality
\[
    \pi^A (\Phi) = \sup_{\mathbb{Q} \in \mathcal{M}} \sup_{\tau \in \mathcal{T}(\mathbb{F})^m} \mathbb{E}^{\mathbb{Q}} \left[\Phi_{\tau} \right] = \sup_{\overline{\mathbb{Q}} \in \overline{\mathcal{M}}} \mathbb{E}^{\overline{\mathbb{Q}}} \left[ \Phi \right].
\]

Furthermore, placing constraints on how these American options can be exercised gives us more non-trivial examples:
\begin{itemize}
\item Waiting period \(n \in \mathbb{N}\) after an execution; in such cases, our collection of admissible plans \(\mathcal{D} := \left\{ c \in \mathbb{T}^{m}: \|c_i - c_j\| \in [n, N], ~\forall 1 \leq i, j \leq N \right\}\); that is, the buyer has to wait at least \(n\) periods after an execution of one of the American options;
\item Limit of \(m' \leq m\) on the number of options exercised each time; hence, the collection admissible plans \(\mathcal{D} := \left\{ c \in \mathbb{T}^m : \sum_{i = 1}^{m} \mathbf{1}_{c_i = k} \le m', \forall k \in \mathbb{T} \right\}\).
\end{itemize}
Any choice outside \(\mathcal{D}\) would result in the payoff being set to \(-\infty\), as pointed out in \Cref{rm:infty}.

\subsection{Option on options recursively}
\label{sec:org8c5f0c1}
Suppose there is such a contract that allows the buyer to choose until time \(K\) before the maturity date \(N\), from a set of options, 
where each element is an option on another set of options, and so on so forth. 
Hence, we have the set of available options \(\mathcal{A}\) at each time as our action space and \(\mathcal{C} := \mathcal{A}^{K}\) as  the collection of all possible action plans. Then the collection of admissible plans \newline \(\mathcal{D} := \left\{ c \in \mathcal{C}: c_{k} \in f_{k}(\left( c_{i} \right)_{i \le k - 1}), ~k = 0, \dots, K \right\}\), where \(f_{k} : \left( c_{i} \right)_{i \le k - 1} \rightarrow \mathcal{A}\) produces the set of options available at time \(k\), as a result of the buyer's previous choices.

\printbibliography
\end{document}